\documentclass[sigconf, 9pt]{acmart}

\usepackage{subfigure}
\usepackage{enumitem}
\usepackage{multirow}
\usepackage{xspace}
\usepackage{etoolbox}

\makeatletter
\def\@ACM@checkaffil{
    \if@ACM@instpresent\else
    \ClassWarningNoLine{\@classname}{No institution present for an affiliation}%
    \fi
    \if@ACM@citypresent\else
    \ClassWarningNoLine{\@classname}{No city present for an affiliation}%
    \fi
    \if@ACM@countrypresent\else
        \ClassWarningNoLine{\@classname}{No country present for an affiliation}%
    \fi
}
\makeatother
\setcopyright{none}
\renewcommand\footnotetextcopyrightpermission[1]{}

\newcommand{\ourmethod}{\textit{MIRO}\xspace}

\newenvironment{sitemize}
  {\begin{itemize}[
      label=\textbullet,
      labelindent=0pt,  
      leftmargin=*,     
      labelsep=0.5em,   
      itemsep=0.3em,
      topsep=0pt
  ]}
  {\end{itemize}}

\begin{document}

\title{\ourmethod : Multi-radar Identity and Ranging for Occupational Safety}

\author{Tirthankar Halder}
\authornote{Both authors contributed equally in this work.}
\email{tirthankarhalder70@gmail.com}
\affiliation{%
  \institution{IIT Kharagpur, India}
}

\author{Argha Sen}
\authornotemark[1]
\email{arghasen10@gmail.com}
\affiliation{%
   \institution{IIT Kharagpur, India}
}

\author{Swadhin Pradhan}
\email{swapradh@cisco.com}
\affiliation{%
  \institution{Cisco Systems, USA}
}

\author{Rijurekha Sen}
\email{riju@cse.iitd.ac.in}
\affiliation{%
  \institution{IIT Delhi, India}
}

\author{Sandip Chakraborty}
\email{sandipc@cse.iitkgp.ac.in}
\affiliation{%
  \institution{IIT Kharagpur, India}
}
\begin{abstract}
Occupational exposure to airborne particulate matter (PM) poses a severe health risk in open industrial workspaces such as stone-cutting yards. Conventional monitoring solutions such as wearable PM sensors and camera-based tracking are impractical due to discomfort, maintenance issues, and privacy concerns. We present \ourmethod, a privacy-preserving framework that integrates continuous PM sensing with a multi-radar millimeter-wave (mmWave) \textit{re-identification} (re-ID) backbone. A distributed network of PM sensors captures localized pollutant concentrations, while spatially overlapping mmWave radars track and re-associate workers across viewpoints without relying on visual cues. To ensure identity consistency across radars, we introduce a \textit{GAN-based view adaptation network} that compensates for azimuthal distortions in range-Doppler (RD) signatures, combined with \textit{correlation-based cross-radar matching}. In controlled laboratory experiments, our system achieves a \textit{re-ID F1-score of 90.4\%} and a \textit{mean Structural Similarity Index Measure (SSIM) of 0.70} for view adaptation accuracy. Field trials in rural stone-cutting yards further validate the system's robustness, demonstrating reliable worker-specific PM exposure estimation.
\end{abstract}

\maketitle

\section{Introduction}
\sloppy



Occupational safety issues arise from health risks due to the coupling between \textit{what a worker is doing and where one is doing it}. In workspaces, such as construction, mining, stone cutting, and metal fabrication, environmental hazards are highly activity-dependent. If activities involve abrasive contact, like cutting, drilling, grinding, or welding, those emit intense bursts of particulate matter (PM) and fine respirable particles, which can cause chronic obstructive lung diseases, and cardiovascular complications~\cite{akbar2010effectiveness,akbar2007crystalline,halvorsen2025measurements}. Similarly, tasks such as torch cutting or solvent cleaning introduce transient chemical vapors, creating health problems due to exposure to problematic volatile organic compounds (VOCs)~\cite{su2018exposures}. These exposures fluctuate quickly as workers alternate between operations, move across zones, and share confined air-spaces. So, accurate occupational-risk assessment, which is essential in improving the safety of the workspace, requires joint, time-aligned measurement of \textit{both activity and pollutant concentration, not just coarse ambient averages}.

Despite advances in industrial IoT and safety analytics, such fine-grained exposure monitoring remains rare in outdoor and semi-outdoor workplaces, especially in developing regions~\cite{johannessen2020embedded,zhou2018examining}. Cameras and wearable sensors, the two dominant approaches, each face fundamental limitations. Cameras provide rich behavioral context but raise severe privacy and compliance concerns, often leading to worker resistance~\cite{nord2006monitoring}, which is further validated by the survey done in our own experiments (see \figurename~\ref{fig:survey}). Wearable PM and motion sensors, while privacy-safe, are intrusive, require maintenance, and bias behavior under heavy workloads~\cite{hansel2018put}. So, occupational-risk monitoring for ensuring safety needs a modality that is simultaneously continuous, privacy-preserving, and maintenance-free. 

Millimeter-wave (mmWave) radars fill this need for an infrastructure-embedded, device-free seamless sensing method~\cite{sen2024continuous, zhou2025mmmulti, feng2022wi}. Operating in the $30-300$ GHz range with multi-gigahertz bandwidth, mmWave can capture centimeter-level range resolution and micro-Doppler signatures~\cite{wang2024rdgait,liu2024pmtrack,sen2024continuous}. These signatures are fine-grained modulations caused by limb movements and gait cycles. Because of its high spatial resolution, mmWave is an ideal candidate for monitoring complex environments. Furthermore, the short wavelength (1-10\,mm) is orders of magnitude larger than airborne dust particles. It makes propagation essentially immune to scattering and attenuation from particulates~\cite{abuhdima2021effect}. Moreover, mmWave's small antenna sizes enable \textit{compact, low-profile units} that can be integrated into poles or walls~\cite{kong2024survey}. Beyond these resolution and robustness advantages, mmWave sensing also provides a distinct opportunity for \textit{movement-pattern-based identification}. Unlike vision-based systems, where appearance cues such as clothing or background context can vary drastically across viewpoints, mmWave sensing depends primarily on coarse grained motion-induced micro-Doppler and range dynamics information. As a result, \textit{identity signatures in mmWave remain largely invariant} to background and viewpoint, provided that the underlying movement pattern is consistent.

\begin{figure}
    \centering
    \includegraphics[width=0.35\textwidth]{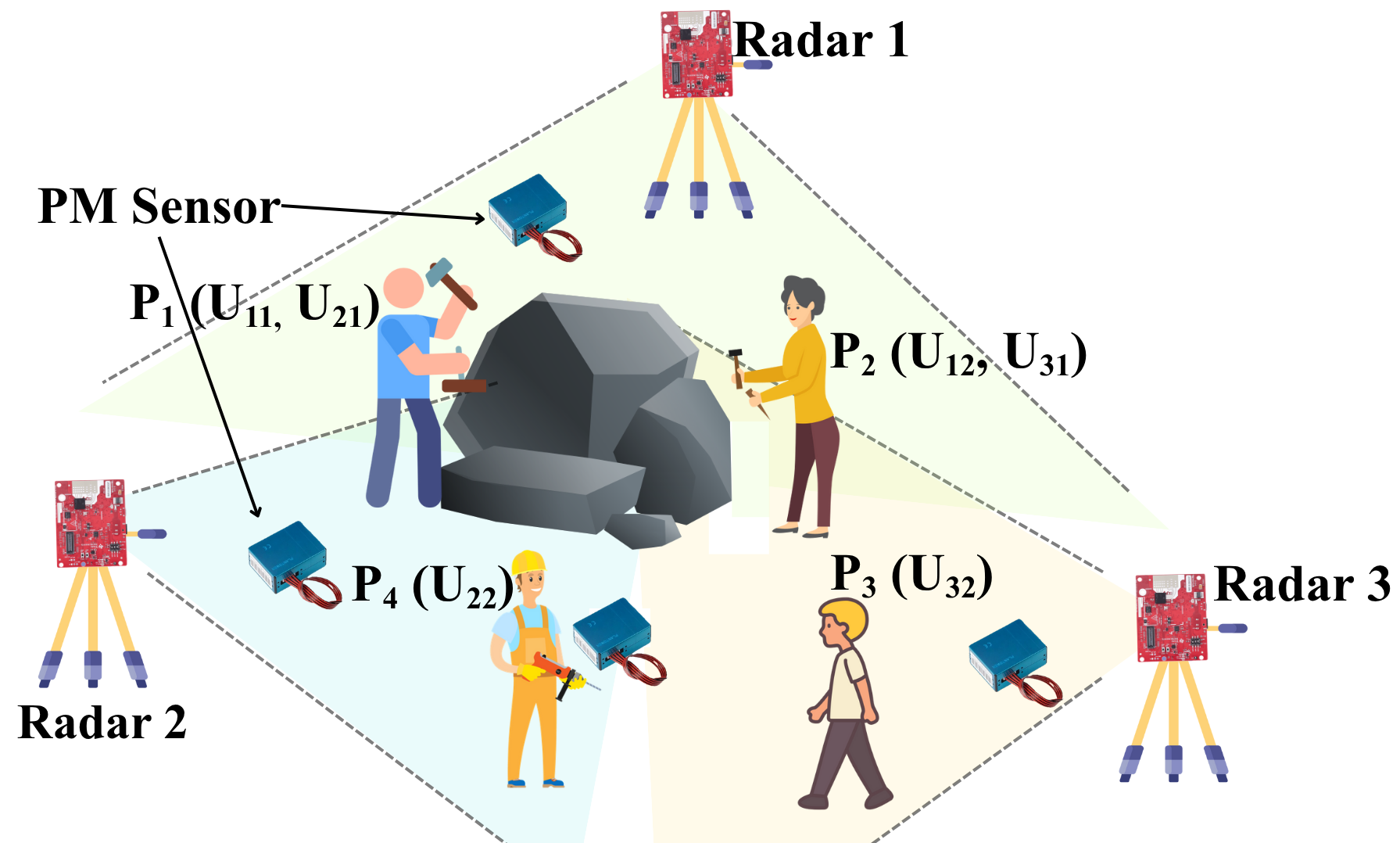}
    \caption{\ourmethod's multi-mmwave radar re-identification (re-ID) in action. It resolves cross-view identity ambiguity due to overlapping regions, where each radar assigns distinct local user IDs (e.g., $U_{11}$, $U_{21}$) to the same worker ($P_1$), by performing activity-correlated re-ID, while co-located PM sensors enable personalized exposure estimation.}
    \label{fig:overview}
    \vspace{-0.4cm}
\end{figure}
Yet scaling mmWave from controlled, single-sensor experiments to \textit{large, real-world environments} introduces a fundamental systems challenge: \textbf{multi-radar re-ID}. A single radar's coverage area is limited~\cite{iwr1843boost,iovescu2020fundamentals}; monitoring large or irregularly shaped spaces requires deploying multiple radars with overlapping fields of view. As shown in \figurename~\ref{fig:overview}, such overlapping coverage introduces cross-view identity ambiguity, where the same worker may be assigned different local user IDs by separate radars observing from distinct viewpoints. This creates the need to consistently link detections of the same individual across sensors. Unlike camera-based re-ID, there are no visual appearance cues; unlike Wi-Fi or UWB~\cite{ren2022person,liu2020vision,garg2025large,mohammadmoradi2017room}, there are no persistent MAC addresses or tags~\cite{kilic2013device}. Without \textit{robust identity persistence} across sensors, trajectories fragment where workers usually operate within semi-static task zones~\cite{papaioannou2016tracking}. 
\subsection{Closing the Multi-Radar re-ID Gap}
Current mmWave re-ID methods typically operate under controlled conditions or rely on trajectory stitching~\cite{han2025mmreid, liu2024mission,wang2025user}. In semi-static industrial environments, workers often remain within fixed task areas, yet these approaches still struggle to maintain reliable identification. The micro-Doppler signatures serve as cues, but they remain sensitive to \textit{changes in viewpoints} and \textit{variability in motion}. 


\begin{figure}[t]
  \centering
  \includegraphics[width=0.30\textwidth]{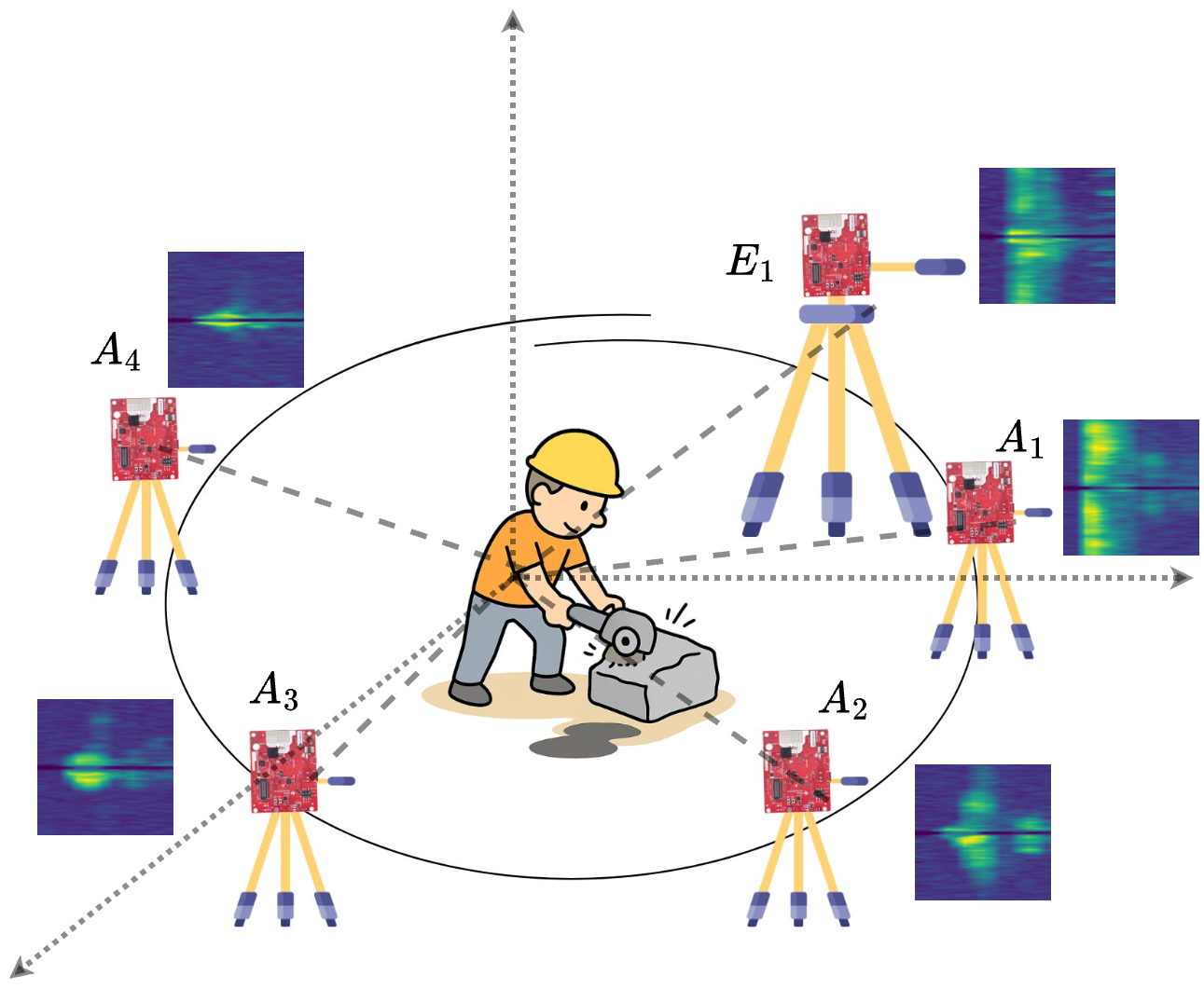}
  \caption{
  Illustration of azimuthal dependence of range-Doppler (RD) signatures. In stone-work, most motions (body/tool) occur within the azimuthal-plane, producing variations in azimuth angles ($A_i$). Elevation-plane motions ($E_i$) contribute less, yielding nearly identical RD responses.}
  \label{fig:azimuth_dependence}\vspace{-0.4cm}
\end{figure}

To address this, we design a \textit{Pix2Pix-based micro-Doppler view adaptation network} tailored for multi-radar operation. Each radar produces time-synchronized 3D point clouds along with RD heatmaps for all detected users. To ensure robust per-radar user localization, we develop a clustering framework called \textit{TDSCAN} ( Temporal Doppler Spatial Clustering ) that integrates Doppler-based filtering with temporal consistency matching (\S\ref{sec:method}). Then, the Pix2Pix-based \textit{view adaptation module} (\S\ref{sec:method}) learns a supervised mapping between paired RD signatures captured from different radar viewpoints. This translation compensates for angle-dependent distortions in the \textit{azimuth plane} of our target workspace.

In particular, the majority of motion in stone-cutting environments such as cutting, grinding, and chipping occurs predominantly within the azimuthal plane, while elevation-plane motion remains minimal, as shown in \figurename~\ref{fig:azimuth_dependence}. As a result, viewpoint variation primarily manifests as azimuthal shifts in the RD signature rather than significant elevation distortions. By modeling these azimuth-dependent transformations, the network captures the essential geometry of motion without overfitting to environment- or activity-specific nuances. Consequently, when the adaptation focuses solely on motion patterns intrinsic to the activity rather than on background clutter, the learned representation generalizes across unseen workspaces and task types. 
\subsection{Real-World Validation in Harsh Settings}
To comprehensively assess the generalizability of \ourmethod, we conducted a two-phase evaluation in both controlled laboratory experiments and real-world field deployments in diverse stone-processing environments. This multi-stage validation enables us to isolate the performance of the view-adaptation and re-ID modules under ideal conditions and then assess their resilience to environmental variability, such as, dust exposure, and multipath effects.

In the controlled laboratory setup, we recruited $3$ participants and instructed them to emulate representative stone-working activities, including chipping, grinding, and polishing gestures. Two mmWave radar nodes were positioned at different azimuthal viewpoints to capture paired RD data simultaneously for each activity. The participants’ ground-truth spatial trajectories were recorded using a Vicon Tracker 4.0 motion capture system with five infrared cameras operating at 100~Hz, providing sub-millimeter accuracy. This controlled setup allowed precise evaluation of the view-adaptation and cross-radar identity consistency modules. The paired multi-view data collected in this phase also used in training our Pix2Pix-based view transformation model.

To evaluate performance under real-world conditions, we deployed \ourmethod~across three distinct stone-processing environments, each presenting unique spatial and operational characteristics. The first site, a large open-air marble processing facility, involved multiple workers simultaneously performing cutting, chipping, grinding, and polishing operations on heavy stone blocks. This site featured continuous dust emission, overlapping worker trajectories, and wide radar separation, testing both dust immunity and cross-node identity persistence. The second site, a semi-mechanized stone-cutting factory, was dominated by high-speed cutting machines generating dense airborne particulates and strong mechanical Doppler signatures. This setting tested the system's ability to maintain accurate worker identification in the presence of intense background motion. The third deployment took place in an indoor construction workspace where marble slabs were being shaped and installed.
\subsection{Contributions}
In summary, this paper makes the following key contributions:

\noindent \textbf{1. \ourmethod~Framework.}  
We present \ourmethod, a multi-radar re-ID framework built on top of robust multi-radar localization enabled by our \textit{TDSCAN} clustering module (\S\ref{sec:method}).
\ourmethod~introduces an \emph{azimuth-aware Pix2Pix view-adaptation network} that translates RD signatures across radar viewpoints. 
Our model learns viewpoint transformations in \emph{azimuthal motion geometry}, and adapts across unseen workspaces and activity types. 

\noindent \textbf{2. Cross-Radar Association and Identity Persistence.}  
Beyond view adaptation, \ourmethod{} incorporates a \emph{cross-radar association engine} that computes \emph{spatio-temporal correlation scores} between view-adapted and observed RD maps, enabling reliable identity propagation across sensing zones and through occlusions. The resulting association graph ensures \emph{global identity consistency} across all radar nodes. In lab tests, \ourmethod{} achieves a re-ID F1-score of up to $90.4\%$ and a mean SSIM of $0.70$.


\noindent \textbf{3. In-the-wild Evaluation.}  
 \ourmethod~was deployed in: (i) an open-air marble facility testing dust immunity and identity persistence, (ii) a semi-mechanized factory dominated by mechanical Doppler interference, and (iii) an indoor workspace with reflective surfaces assessing multi-path resilience. 
 Together, these experiments validate \ourmethod's \emph{robustness} from controlled conditions to complex, real-world industrial settings.

\section{Related Work}
\sloppy

\subsection{Occupational Exposure Monitoring}
As occupational environments become increasingly instrumented through the Industrial Internet of Things (IIoT), numerous embedded sensing systems have been introduced to enable affordable, real-time monitoring of dust, gases, and volatile organic compounds (VOCs)~\cite{johannessen2020embedded,zhou2018examining}. Recent systems combine low-cost particulate sensors with edge computing (e.g., LoRa/LoRaWAN) to produce spatial maps of pollutant concentrations in urban, manufacturing, and construction environments~\cite{chen2014indoor,alsafery2023sensing,karmakar2024exploring}. However, most deployments treat the workspace as a uniform environment, averaging readings over time or space. Such coarse spatiotemporal sampling fails to capture the short-term exposure spikes caused by transient, task-specific operations, which are key determinants of respiratory and cardiovascular outcomes~\cite{akbar2007crystalline, akbar2010effectiveness}. 

\subsection{Activity and Worker Behavior Monitoring}
To contextualize exposure, prior research has explored activity recognition using cameras, wearable sensors, and other modalities. Vision-based systems can infer detailed worker posture, task type, and safety compliance~\cite{yoshimura2024openpack,abdullah2025isafetybench,dallel2020inhard}, offering rich behavioral information for risk analytics. However, the use of cameras in occupational settings raises privacy concerns, data governance, and regulatory compliance~\cite{nord2006monitoring,huang2023rethinking,browne2020camera, rinta2021low}, often leading to worker resistance and legal restrictions. Wearable IMUs, physiological sensors, and smart PPE have also been employed for motion capture and workload estimation~\cite{xia2020robust,niwarthana2024efficient,han2018smart}. Despite privacy benefits, these devices cause discomfort, and behavioral bias, motivating infrastructure-based, \emph{device-free} sensing that passively infers activity~\cite{sen2024continuous,ding2016device}.

\subsection{mmWave and RF-Based Human Sensing}
mmWave radars have demonstrated powerful capabilities for fine-grained human sensing. Operating in the 30--300~GHz spectrum with multi-gigahertz bandwidths, mmWave radars offer centimeter-level range resolution and the ability to capture subtle micro-Doppler modulations induced by limb movements~\cite{wang2024rdgait, liu2024pmtrack, sen2024continuous}. These signatures have been utilized for gait recognition~\cite{wang2024rdgait}, gesture classification~\cite{feng2022wi}, vital signs~\cite{zhu2025measuring,tang2024bsense}, and occupancy estimation~\cite{ren2022person,li2023bmmw,wei2015mtrack}. Unlike optical sensors, mmWave operates reliably under poor illumination, dust, or smoke, and inherently preserves privacy~\cite{abuhdima2021effect}. However, existing studies often assume single-sensor, laboratory-controlled conditions with unobstructed line-of-sight targets. Extending these methods to multi-radar industrial environments requires addressing challenges of sensor coverage and identity continuity across viewpoints.
\subsection{Multi-Sensor Cross-View Re-Identification}

Cross-view re-ID has been extensively studied in computer vision, where visual appearance cues such as texture, color, and pose are exploited to link subjects across cameras~\cite{shen2015person,meng2019weakly}. In contrast, RF-based sensing lacks explicit visual features and must instead rely on motion dynamics, spatio-temporal trajectories, or statistical similarity of doppler profiles. Early mmWave re-ID methods focused on trajectory stitching or clustering in range-time space~\cite{han2025mmreid, liu2024mission,wang2025user}; however, these approaches are designed for single-radar setups and fail to maintain identity consistency across overlapping fields of view in multi-radar deployments. Wi-Fi~\cite{liu2020vision,ren2022person} and Ultra-Wideband (UWB)~\cite{kilic2013device,sun2025wuloc,li2025medusa} systems provide identity persistence through MAC addresses or tag IDs, yet such mechanisms contradict the device-free design goals central to industrial sensing. Recent efforts have begun exploring quasi-biometric properties of micro-Doppler signatures~\cite{wang2024rdgait,liu2024pmtrack}, but these features exhibit strong viewpoint dependence. As a result, maintaining consistent user identity across multiple mmWave radar viewpoints with overlapping fields of view remains an open challenge~\cite{papaioannou2016tracking}.

View adaptation and domain translation techniques have been highly successful in computer vision for cross-domain mapping and style transfer (e.g., Pix2Pix~\cite{isola2017image}, CycleGAN~\cite{zhu2017unpaired}).  However, the radar and mmWave re-ID literature has not yet converged on a standard approach for direct image-to-image translation of RD representations.  Instead, most prior mmWave/RF methods rely on trajectory stitching, handcrafted or learned view-invariant descriptors, and supervised learning applied in single-radar settings~\cite{han2025mmreid,liu2024mission,wang2025user}.  These approaches work well when sensors operate independently or when coverage is non-overlapping, but they do not explicitly learn cross-view mappings of micro-Doppler structure and therefore struggle to preserve identity consistency when fields of view overlap or when viewpoint shifts substantially. 

\section{Background and Pilot Study}
\sloppy
In this section, we first introduce the fundamentals of mmWave sensing to establish the radar processing concepts used throughout this work. We then shift focus to a specific real-world use case: monitoring occupational health risks in stone-cutting yards. To ground our study, we begin with a survey of workers' perceived challenges in their daily environment, followed by a pilot deployment of pollution sensors. Based on the limitations of pollution-only monitoring and the workers' concerns about privacy, we motivate the use of unobtrusive mmWave radar. Finally, we present our pilot experiments with radar-based activity monitoring.

\subsection{Preliminaries: mmWave Radar Sensing}\label{sec:prelims}
Commercial off-the-shelf mmWave radars typically operate using the FMCW principle~\cite{rao2017introduction}. In this scheme, the radar continuously transmits frequency-modulated chirps, while simultaneously mixing the TX signal with the RX echo from surrounding objects. This mixing process, often referred to as \textit{dechirping}, produces an \textit{Intermediate Frequency} (IF) signal. 

\subsubsection{Range estimation}\label{subsec:range}
The distance between the radar and a target is obtained from the IF signal offset~\cite{rao2017introduction}. This offset, known as the \textit{beat frequency} ($f_b$), results from the round-trip delay $\tau$. If the chirp of duration $T_C$ sweeps over a bandwidth $B$, the chirp slope can be expressed as $S = \frac{B}{T_C} = \frac{f_b}{\tau}$. Since the delay is given by $\tau = \frac{2d}{c}$, where $d$ is the object distance and $c$ is the speed of light, the distance can be calculated as $d = \frac{c}{2} \cdot \frac{T_C}{B} \cdot f_b.$
To estimate $f_b$, a Fast Fourier Transform (FFT), called the \textit{range-FFT}, is applied to the IF signal. The resulting spectrum contains peaks at frequencies corresponding to object reflections, thereby providing the range estimate.

\subsubsection{Velocity estimation}\label{sec:velocity}
To extract velocity information, the radar transmits a sequence of $N$ chirps, each separated by a duration of $T_C$. If a target moves at speed $v$, the motion introduces a phase shift between two consecutive received chirps. This phase shift, $\Delta \phi$, is given by $\Delta \phi = \frac{4\pi v T_C}{\lambda},$ where $\lambda$ is the wavelength of the carrier. By performing a second FFT, referred to as the \textit{Doppler-FFT}, on the phase variations across chirps, the velocity of the target can be estimated. This information is organized into a two-dimensional \textit{range-Doppler} matrix $\mathbb{D}_{D \times R}$, where $D$ and $R$ denote the number of Doppler bins and range bins, respectively.

\subsubsection{Point cloud estimation}\label{sec:pointcloud}
The \textit{point cloud} is extracted from the range-Doppler matrix using the Constant False Alarm Rate (CFAR) detector~\cite{nitzberg1972constant}, which identifies prominent peaks corresponding to actual targets. The resulting point cloud encodes the spatial coordinates $(x_i, y_i, z_i)$, Doppler velocity $(d_i)$, and received signal power $(p_i)$ for each detection. Formally, for $N$ detected objects, the point cloud set $S$ can be expressed as, $S = \bigcup_{i=1}^N \left\{ (x_i, y_i, z_i, d_i, p_i) \right\}$.    
Further details on point cloud generation are available in~\cite{rao2020introduction}.

\subsection{Survey of Stone-Cutting Workers}\label{sec:survey}
To understand the real challenges faced in stone workers' occupational environments, we conducted a survey among laborers in a semi-urban marble processing factory. A total of $8$ workers participated in the survey, with an average age of $43$ years and more than 10 years of work experience in stone cutting. 

\begin{figure}[h]
    \centering
    \subfigure[Occupational risk severity]{
    \includegraphics[width=0.2\textwidth]{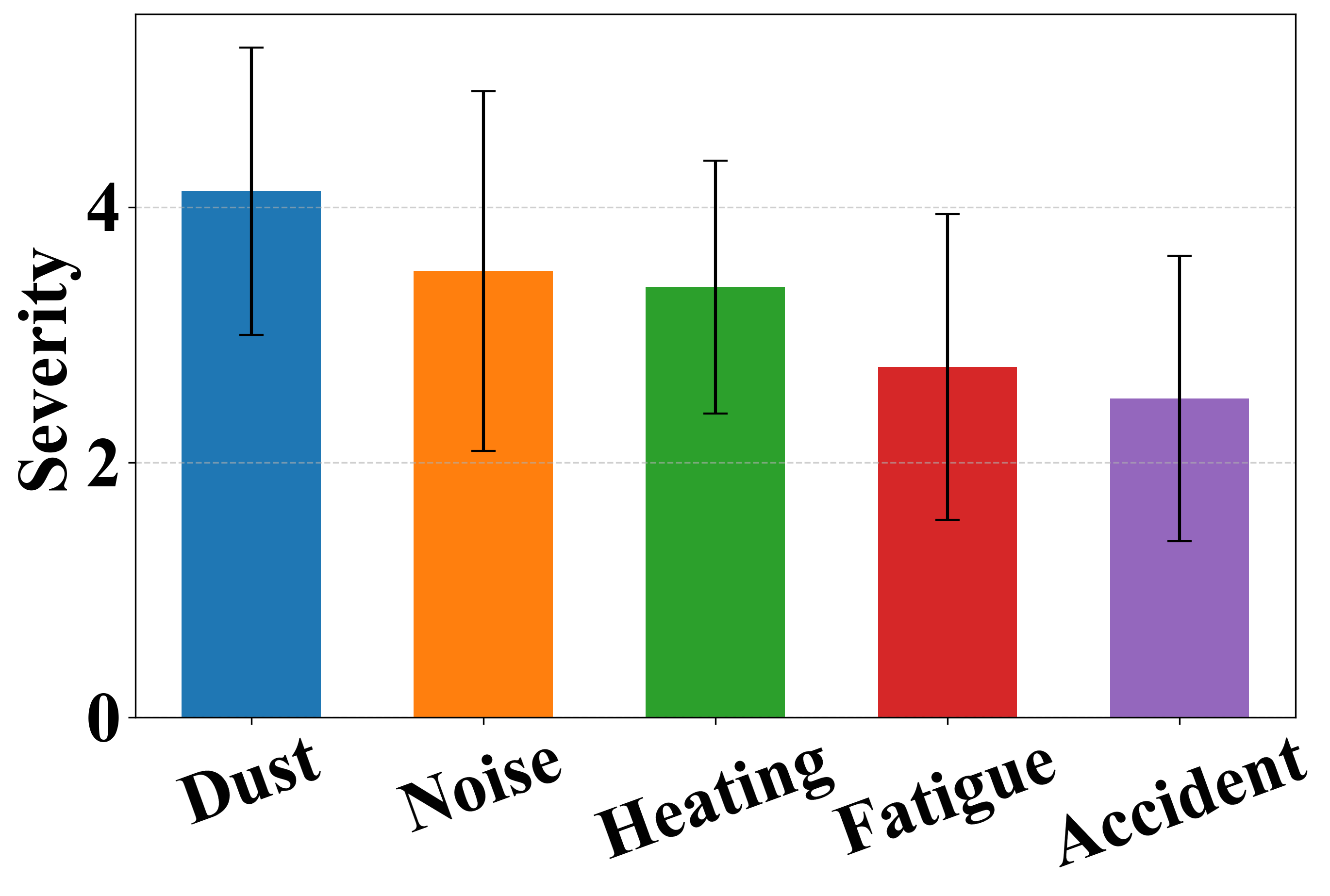}\label{fig:survey1}
    }\hfil
    \subfigure[Modality preferences]{
        \includegraphics[width=0.2\textwidth]{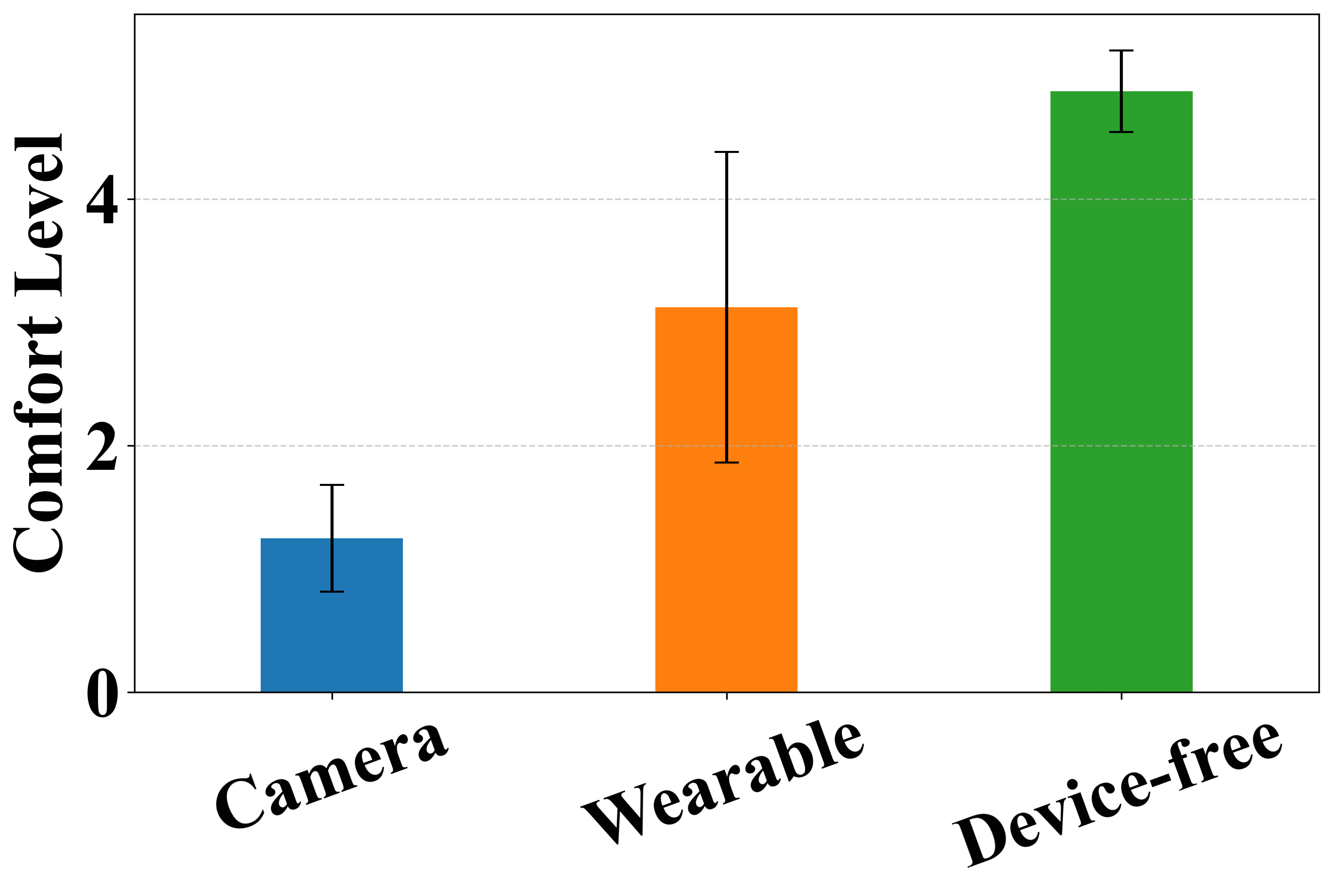}\label{fig:survey2}
    }
        \caption{Survey results from stone-cutting workers.}
    \label{fig:survey}
    \vspace{-0.4cm}
\end{figure}

Workers were then asked to rank the severity of common workplace risks, including (i) dust exposure (respiratory issues, silicosis risk), (ii) noise exposure, (iii) heat and harsh environmental conditions, (iv) repetitive strain and physical fatigue, and (v) risk of accidents from heavy machinery or tools. The results, summarized in \figurename~\ref{fig:survey1}, clearly indicate that dust exposure is considered the most severe challenge, with an average severity score of 4.125 out of 5. Noise and heat were perceived as moderate risks, while repetitive strain and accident risk received relatively lower average scores. This confirms dust exposure as a dominant occupational hazard, emphasizing the need for localized pollution monitoring.

\subsection{Pilot Deployment with Pollution Sensors}
\begin{figure}[h]
    \centering
    \subfigure[Sensor node]{
        \includegraphics[width=0.3\textwidth]{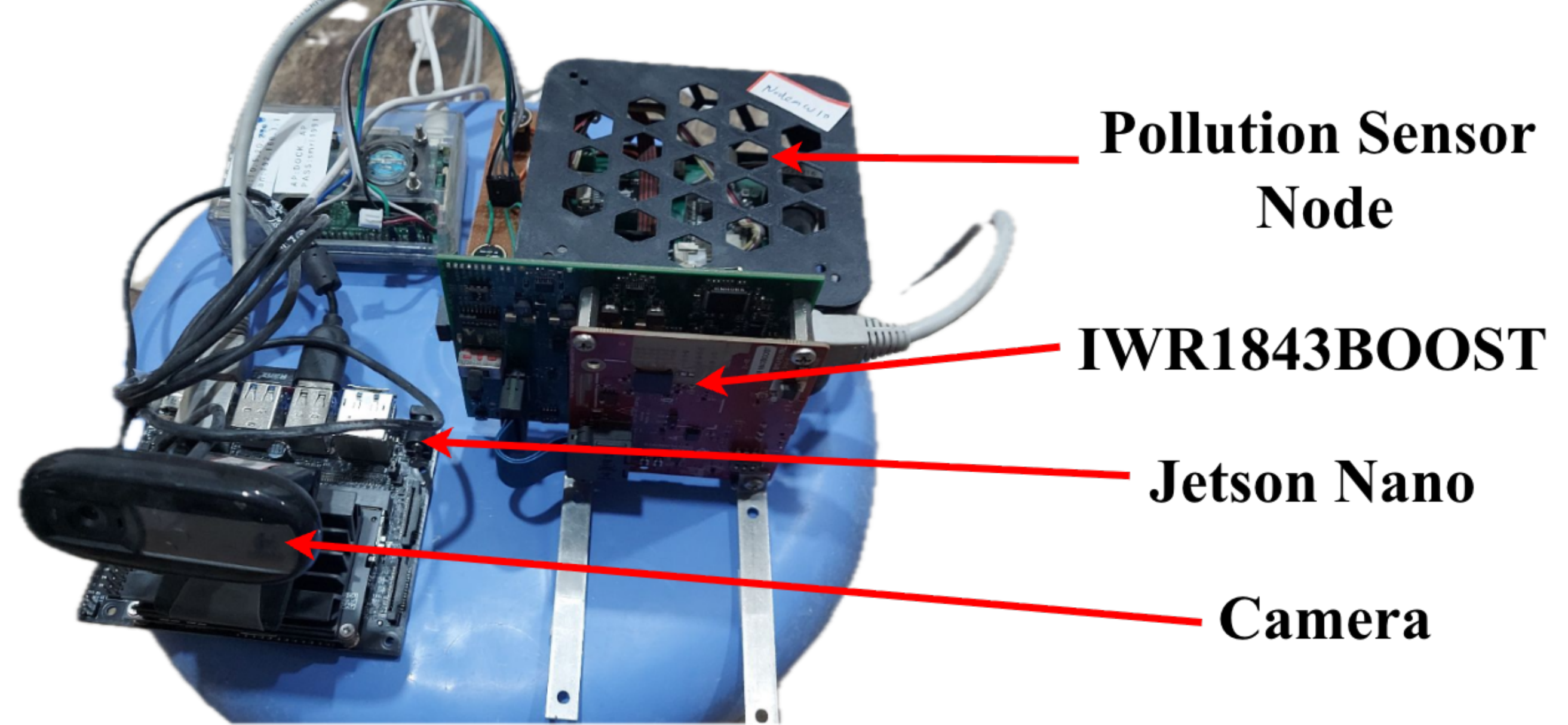}
        \label{fig:node}
    }\\
    \subfigure[Real-world deployment]{
        \includegraphics[width=0.2\textwidth]{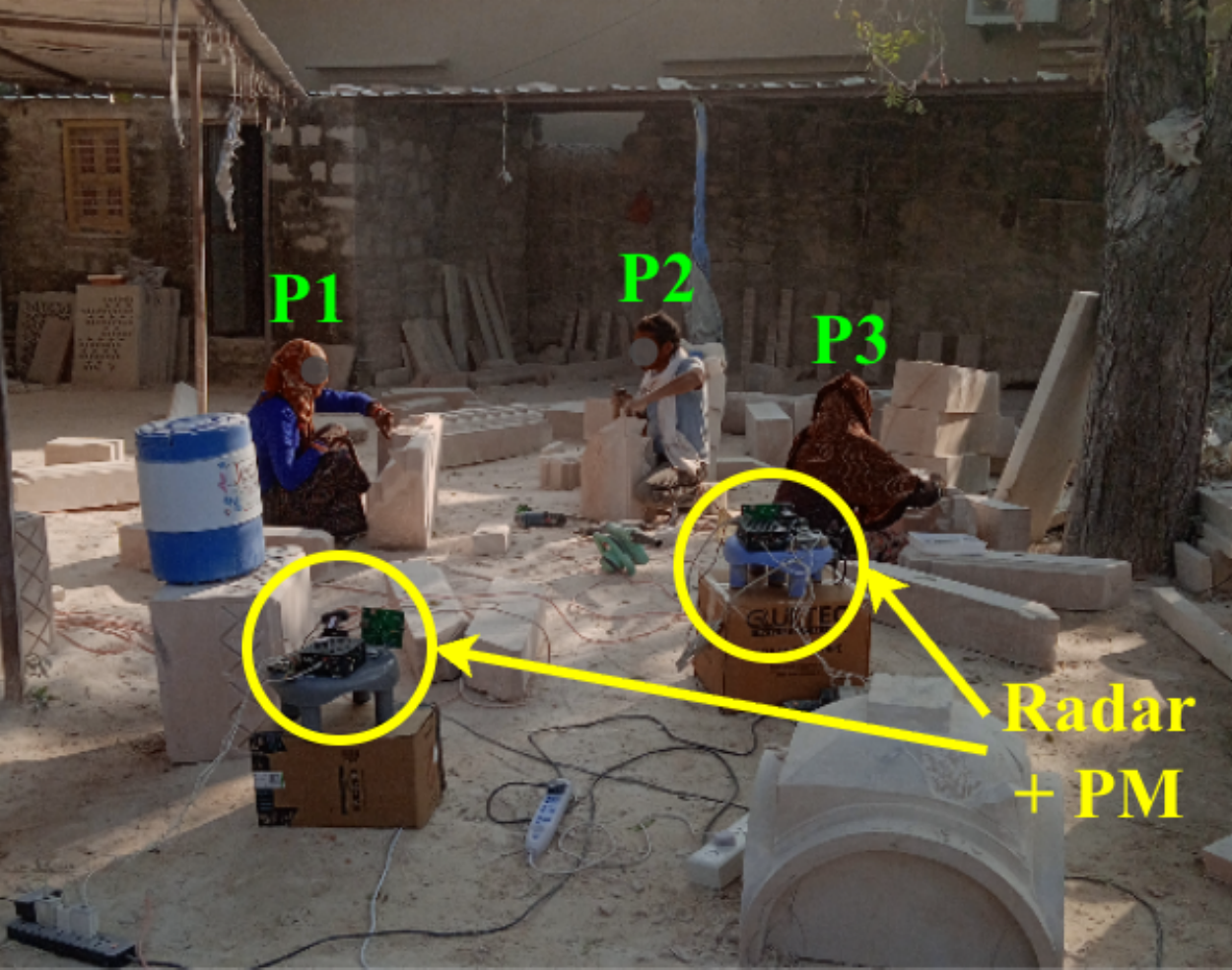}
        \label{fig:deployment_real}
    }\hfil
    \subfigure[Deployment map]{
        \includegraphics[width=0.16\textwidth]{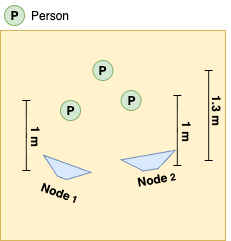}
        \label{fig:deployment_drawio}
    }
    \caption{Detailed experimental setup.}\vspace{-0.4cm}
\end{figure}

Motivated by these findings, we first deployed \textit{DALTON} PM sensors~\cite{karmakar2024exploring} to capture localized pollution levels in a large open-air marbel processing facility. Each sensing node, shown in \figurename~\ref{fig:node}, was capable of measuring $PM_{1.5}$, $PM_2$, and $PM_{10}$ concentrations, along with other gases such as CO, CO$_2$, NO$_2$, and VOCs. The real-world deployment of these nodes in the yard is illustrated in \figurename~\ref{fig:deployment_real}, while a schematic layout of the deployment is provided in \figurename~\ref{fig:deployment_drawio}. In this pilot study we primarily focus on PM$_x$ measurements, which have direct links to occupational hazards such as silicosis~\cite{akbar2007crystalline}.

\subsubsection{PM Exposure Analysis Across Activities} 
\begin{figure}[!h]
    \centering
    \subfigure[Comparison of activities]{
        \includegraphics[width=0.22\textwidth]{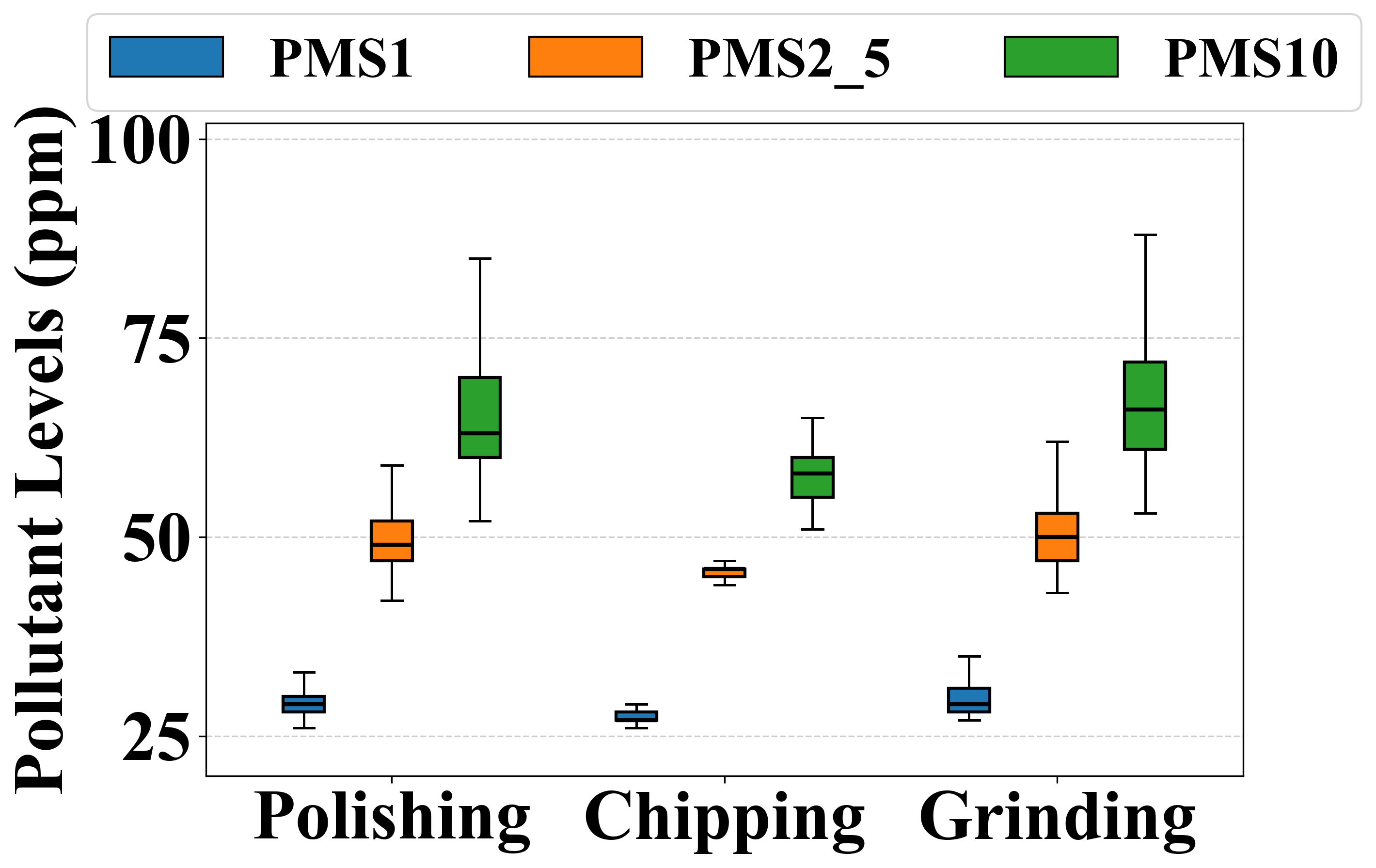}
        \label{fig:polActivity}
    }
    \subfigure[Impact of sensor proximity to source ]{
        \includegraphics[width=0.22\textwidth]{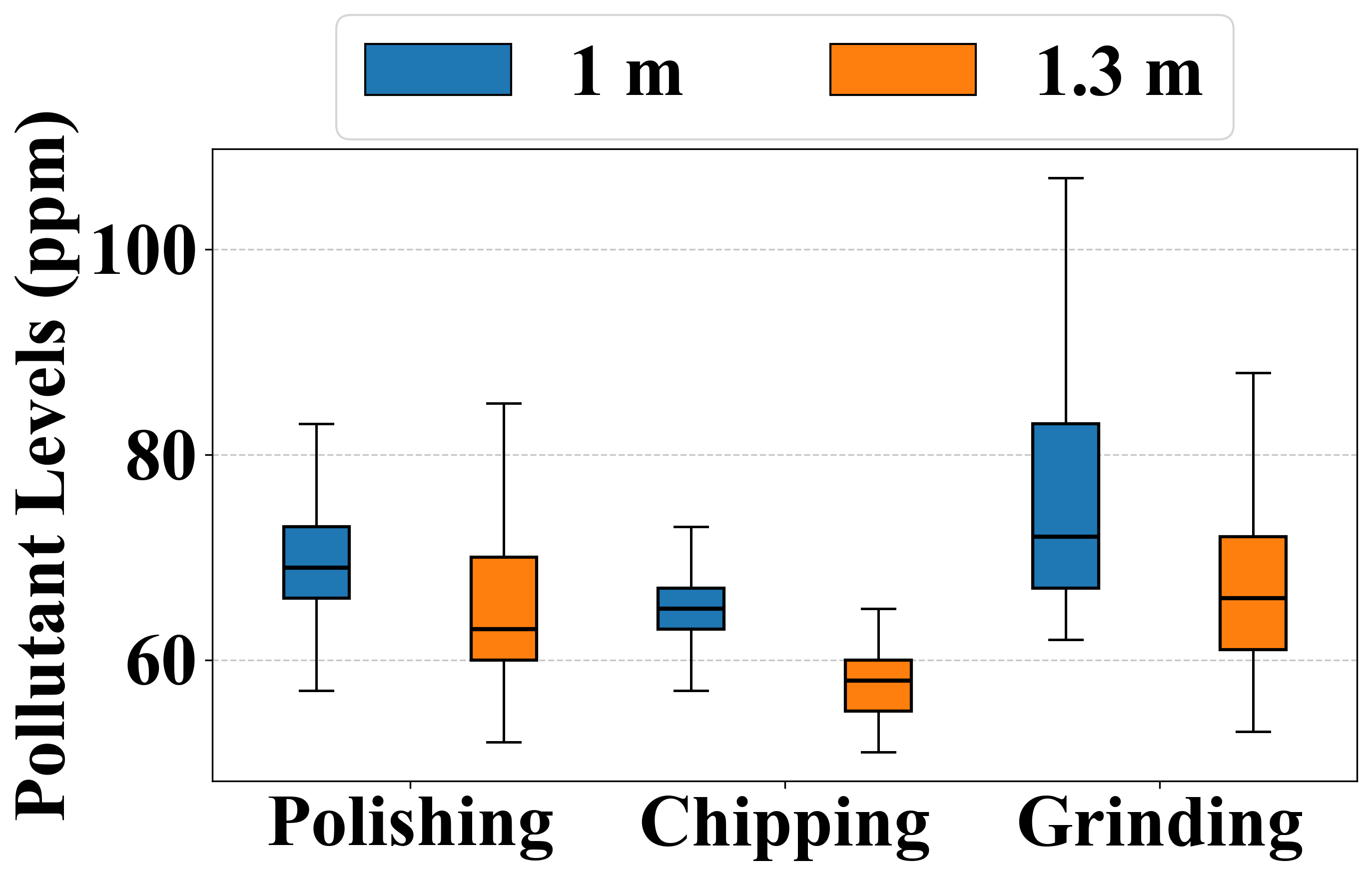}
        \label{fig:polDistance}
    }
    \caption{PM exposure analysis.}\vspace{-0.4cm}
    \label{fig:pmExposurePilot}
\end{figure}

Stone-cutting tasks are significantly diverse in their dust-generation potential. We measured localized PM concentrations during three common activities: polishing, chipping, and grinding. Our results as shown in \figurename~\ref{fig:polActivity} demonstrate a clear gradient in exposure levels. Grinding generated the highest pollution levels, often producing PM concentrations that exceed occupational safety thresholds by orders of magnitude~\cite{pmaqi2017}. 
The fine particles generated during grinding remained suspended longer and spread further. Polishing exhibited moderate PM levels, typically lower than grinding, but still significantly above background levels. Chipping produced the least amount of PM among the activities.    

Thus pollutants' exposure is not uniform across the workspace but is strongly tied to the nature of the activities being performed. Thus continuous task or activity recognition is crucial in this direction to have activity-specific insights. 

\subsubsection{Impact of Sensor Proximity to Dust Source} 
Another key observation from our pilot deployment is the strong effect of distance between the PM sensor and the pollutant source. As shown in \figurename~\ref{fig:deployment_real} we have stone cutting activities happening at different distances with respect to the pollutant sensors ($\approx$ 1 m and 1.3 m as shown in \figurename~\ref{fig:deployment_drawio}). Sensors placed closer to the grinding activity recorded peak concentrations several times higher than those located only a few meters away (see \figurename~\ref{fig:polDistance}). This distance sensitivity implies that sparse placement of PM sensors can lead to biased exposure estimates if worker proximity is not tracked. 

Thus, it becomes essential to link worker trajectories with spatially varying PM measurements to estimate individualized exposures. While PM sensors provide valuable measurements of localized pollutant concentrations, they suffer from several limitations that prevent their use as a standalone solution. First, in high-dust environments such as stone-cutting yards, PM sensors can clog or saturate, leading to unreliable readings. Second, their measurements are highly sensitive to placement and orientation: a sensor located only a few meters away from the activity can record drastically lower exposure than one placed nearby. In addition, pollution levels alone cannot tell which worker is generating dust or which activity is responsible, making worker-specific exposure insights impossible. Due to these limitations, alternative monitoring approaches are needed that link exposure levels to worker positions and activities. Among the common options are camera-based monitoring and wearable sensors, each with its own privacy, comfort, and reliability trade-offs. To better understand worker perceptions of these modalities, we conducted a survey on monitoring preferences.

\subsection{Constraints on Monitoring Modalities}
We asked workers about their comfort levels with different monitoring approaches. The survey results, summarized in \figurename~\ref{fig:survey2}, indicate that \textit{camera-based monitoring} was generally viewed as intrusive and raising strong privacy concerns, while \textit{wearable-based monitoring} (e.g., smart bands) was reported as uncomfortable or impractical given the physical demands of stone-cutting. In contrast, workers expressed much higher acceptance for device-free monitoring solutions that do not rely on either cameras or wearables.

These findings suggest that mmWave radar can be a promising direction: it offers device-free, privacy-preserving robust sensing in dusty environment. However, a single mmWave radar cannot provide complete coverage due to limited FoV and frequent occlusions. This limitation motivates the exploration of multi-radar setups and also introduces new challenges in making activity recognition consistent across different viewpoints.

\begin{figure}[h]
    \centering
    \subfigure[Range-Doppler signatures of activities]{
    \includegraphics[width=0.45\textwidth]{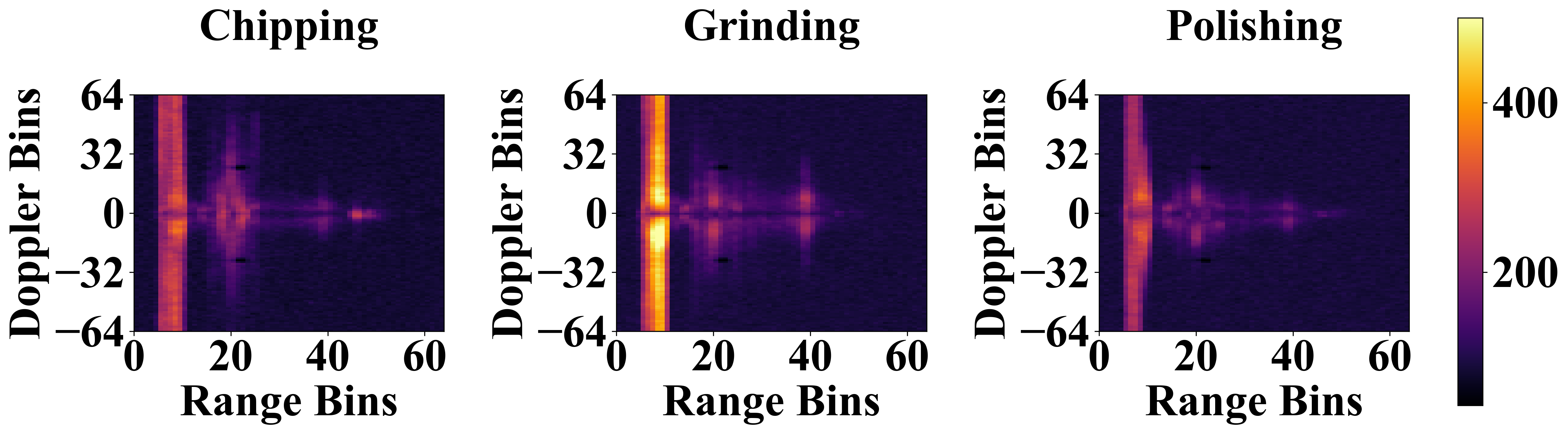}
    \label{fig:activitySpectrograms}
    }\hfil
    \subfigure[t-SNE feature distribution]{
        \includegraphics[width=0.24\textwidth]{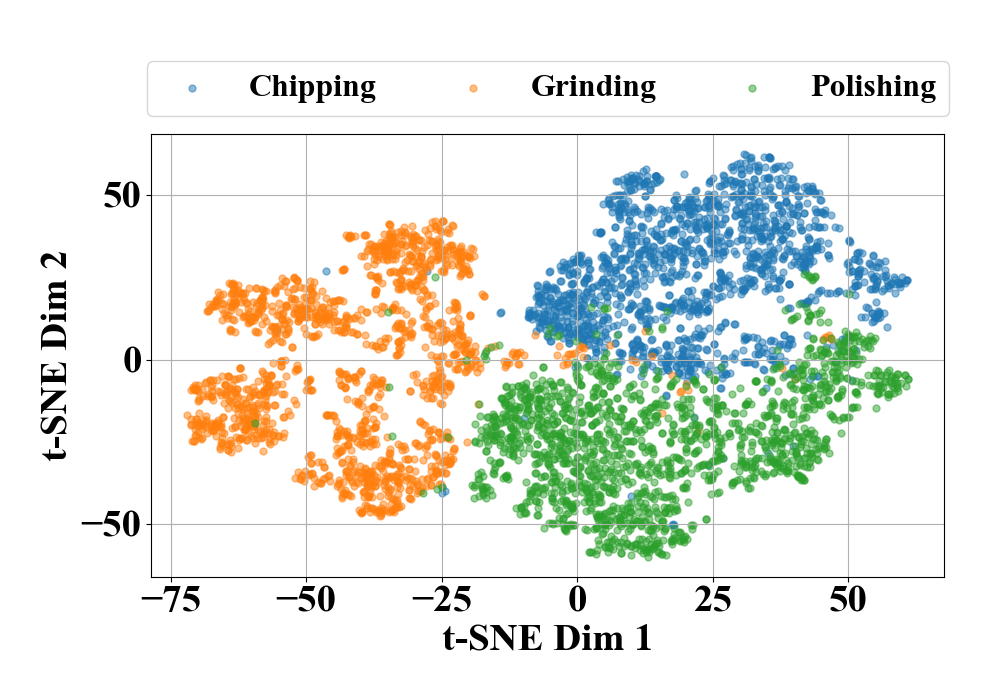}
        \label{fig:tsneActivity}
    }\hfil
    \subfigure[Range-dependent analysis]{
        \includegraphics[width=0.20\textwidth]{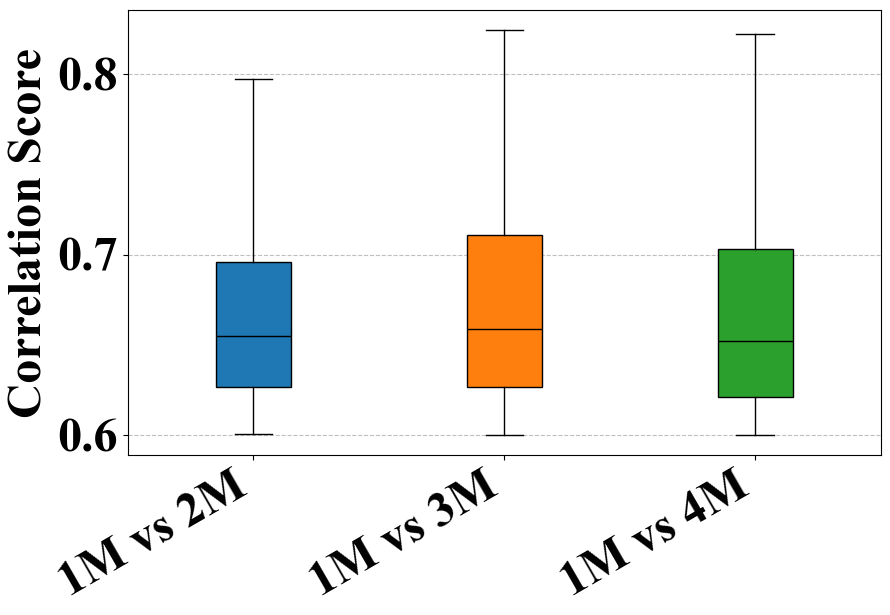}\label{fig:chippingRange}
    }\\
    \subfigure[Correlation of signatures across angular viewpoints]{
        \includegraphics[width=0.4\textwidth]{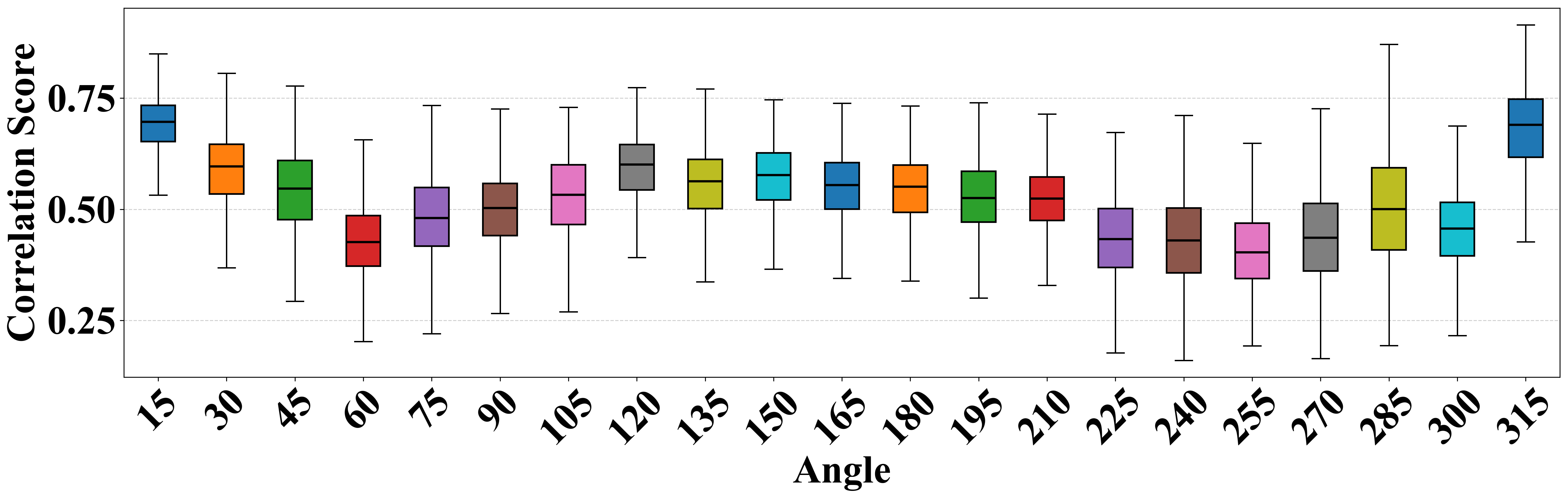}\label{fig:polishingangles}
    }
    \caption{Radar-based activity analysis.}\vspace{-0.5cm}
\end{figure}

\subsection{Activity Signature Analysis Using mmWave}
Even though PM sensors measure localized concentrations, they cannot differentiate between workers or activities that generate pollution. To address this, we leverage mmWave radar's micro-Doppler signatures. As shown in \figurename~\ref{fig:activitySpectrograms}, different activities produce distinct range-Doppler spectrograms, enabling the possibility of activity classification and separation of workers' roles. To highlight this, we apply t-distributed Stochastic Neighbor Embedding (t-SNE)~\cite{maaten2008visualizing} to project the features from raw range-Doppler data into a two-dimensional feature space. As shown in \figurename~\ref{fig:tsneActivity}, the embeddings for different activities form distinct clusters, demonstrating that micro-Doppler features can classify tasks.

However, this discriminability is not invariant to sensing geometry. When the same activity is observed from different angular viewpoints, the resulting spectrograms exhibit reduced correlation compared to a reference view taken at $0^\circ$ along the radar boresight. For instance, grinding observed laterally (at $90^\circ$) produces a substantially different Doppler spread than when observed head-on, as shown in \figurename~\ref{fig:polishingangles}. In contrast, changes in range have a weaker effect: signatures recorded at different distances but similar angles maintain higher correlation, as shown in \figurename~\ref{fig:chippingRange}.

\begin{figure*}
    \centering
    \includegraphics[width=0.8\textwidth]{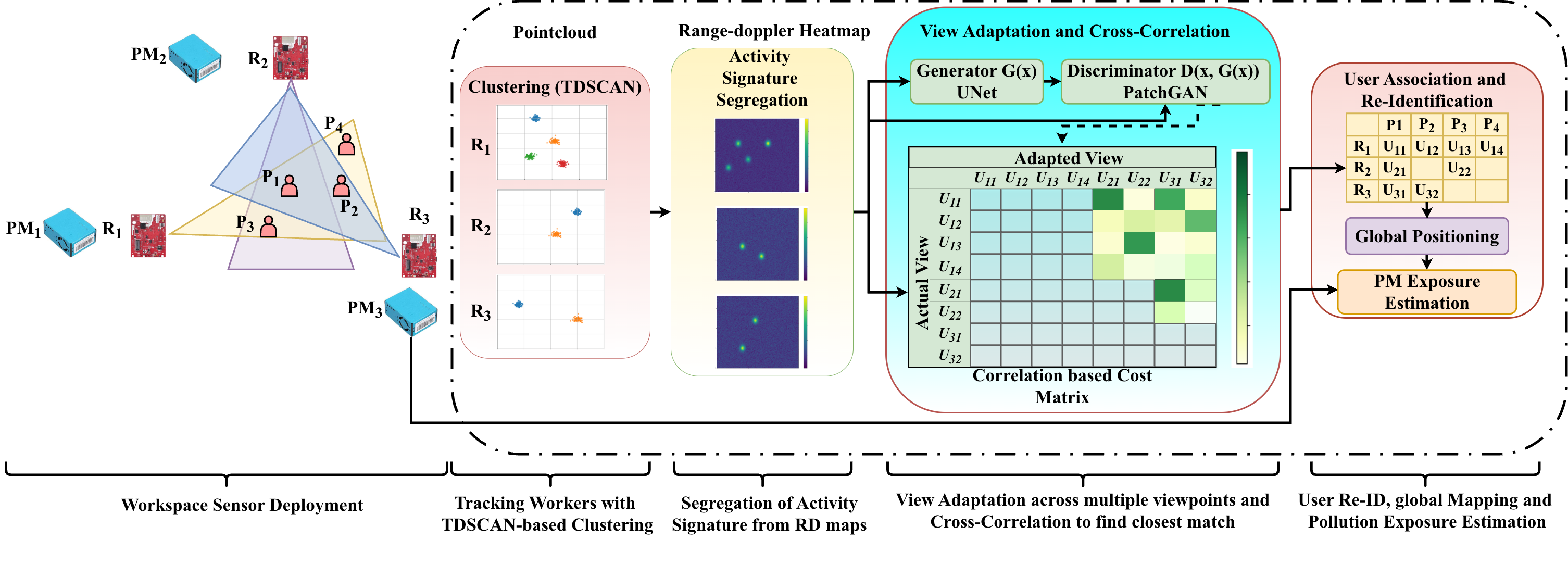}
    \caption{System overview diagram of \ourmethod.}
    \label{fig:stoneworkersbd}
    \vspace{-0.4cm}
\end{figure*}

These findings highlight that although multiple sensors are required for full coverage, their differing viewpoints introduce distortions that can confuse recognition models. This motivates the design of a view-adaptation mechanism capable of aligning micro-Doppler features across arbitrary angles, thereby ensuring reliable activity discrimination in heterogeneous, multi-radar environments.

\section{Methodology}\label{sec:method}
\sloppy
We aim to achieve robust multi-radar user re-ID and personalized pollutant exposure estimation in the workplace. To accomplish this, we combine radar-based localization, activity signature extraction, multi-radar view adaptation, identity correlation, and pollution mapping with global positioning (shown in \figurename~\ref{fig:stoneworkersbd}).
\subsection{User Localization via Pointcloud Clustering}
A multi-radar environment makes accurate user localization difficult because each radar observes semi-static subjects from different viewpoints and under varied clutter and signal sparsity. Conventional clustering algorithms such as DBSCAN~\cite{ester1996density} are ineffective because radar detections are sparse, non-uniform, and temporally inconsistent across sensors. The \textit{fixed density} and \textit{neighborhood thresholds} used by DBSCAN make it sensitive to these variations, often causing fragmented clusters for weak detections or merged clusters when reflections from multiple users overlap. Such inconsistencies are further amplified in a multi-radar setup, where small clustering errors at each radar propagate across viewpoints, leading to misaligned user representations and degraded performance in downstream view adaptation and cross-radar identity matching. To overcome these limitations, we propose \textbf{TDSCAN (Temporal Doppler Spatial Clustering)}, a modified clustering framework that integrates \emph{Doppler-based filtering} with \emph{temporal consistency matching} to ensure reliable localization across multiple radar nodes in cluttered industrial environments.


The key motivation behind TDSCAN is two-fold:  
(1) \textbf{Doppler-based filtering} removes static clutter points whose Doppler values are near zero, isolating reflections corresponding to slow human motions. Since workers in our scenario perform quasi-static activities (e.g., grinding and not walking/heavy exercises involving significant body movements), valid reflections exhibit small but non-zero Doppler velocities. Thus, we apply a lower and upper Doppler bound $(\tau_{\min}, \tau_{\max})$ to retain only those points satisfying $\tau_{\min} < |v_i| < \tau_{\max}$. In our implementation, thresholds of $\tau_{\min}=0.05$~m/s and $\tau_{\max}=1.0$~m/s effectively distinguish human workers from non-human static clutter and high-velocity noise.  
(2) \textbf{Temporal consistency matching} enforces continuity of detected clusters across consecutive frames. Rather than treating each frame independently, TDSCAN accumulates detections over a short temporal window (10 frames, or approximately 1~second at 10~FPS), then performs centroid-based track association using a distance-constrained assignment (Hungarian algorithm). This temporal linkage mitigates discontinuities in user localization and suppresses false positives from transient reflections. Formally, after Doppler-based filtering, we obtain a reduced set of points
\begin{equation}
\mathcal{P}' = \{p_i = (x_i, y_i, z_i, v_i) \in \mathcal{P} \;|\; \tau_{\min} < |v_i| < \tau_{\max}\}.
\end{equation}


TDSCAN then operates on the Doppler-filtered point set $\mathcal{P}'$, aggregating spatially and temporally coherent points into user clusters 
A point $p_i$ is a \emph{core point} if:
\begin{equation}\label{eq:2}
    |\{p_j \in \mathcal{P}' \;|\; \|p_i - p_j\| \leq \epsilon\}| \geq minPts,
\end{equation}
where $\epsilon$ is the neighborhood radius and $minPts$ is the minimum number of neighboring points. Each cluster $\mathcal{C}_k$ represents a user track, with its centroid giving the instantaneous 2D position.
\begin{equation}\label{eq:3}
    \mathbf{u_k} = \frac{1}{|\mathcal{C}_k|}\sum_{p_i \in \mathcal{C}_k} (x_i, y_i).
\end{equation}

To maintain temporal coherence, cluster centroids from consecutive windows are matched based on spatial proximity. Tracks are updated when a new centroid lies within a predefined distance from an existing track; otherwise, new IDs are initialized. Clusters that appear inconsistently or for fewer than a threshold number of frames are discarded, eliminating short-lived clutter responses. So \textit{TDSCAN} integrates (i) Doppler-aware point selection, (ii) spatial density-based clustering, and (iii) temporal centroid tracking, producing robust, low-latency user localization. 

\subsection{Segregation of Activity Signatures}
Beyond localization, we require discriminative features for re-ID. For each of the localized clusters, we extract the RD heatmaps centered at the user's range bin. Specifically, if the user is localized at range bin $r_0$, we extract RD patches spanning $r \in [r_0-10, r_0 + 10]$ bins over the range axis. This yields a heatmap $\mathbf{H}_k(r,v)$ encoding the RD signature of user $k$. Let $\mathbf{R}(r,v)$ denote the full RD heatmap, then the user-specific activity signature is:
\begin{equation}
    \mathbf{H}_k = \mathbf{R}(r,v)\big|_{r \in [r_0-10, r_0+10]}.
\end{equation}
This selective extraction focuses on isolating motion-induced Doppler components associated with the user's activity while excluding static reflections and background clutter present in other range bins. Since environmental reflections, machinery, and stationary objects contribute primarily to static or near-zero Doppler components, cropping around the user's localized range bin effectively suppresses these environment-dependent features. The resulting $\mathbf{H}_k$ thus captures only the intrinsic motion dynamics, such as limb oscillations or tool-hand interactions, that define the worker’s activity. By filtering out environmental influence at the feature level, these activity signatures become inherently \textit{environment-agnostic}, and form the input for the view-adaptation module.
\subsection{View-Adaptation via Generative Conversion}\label{sec:view_adaptation}
Micro-Doppler signatures are highly viewpoint-dependent — the same activity (e.g., grinding or polishing) produces significantly different RD spectrograms when observed from different radar angles. Such angular variations arise from differences in radial velocity components and relative motion geometry, which distort Doppler spread and intensity patterns. Consequently, direct comparison of RD signatures across radars becomes unreliable.

To mitigate this, we employ a generative view-adaptation module that learns to translate RD signatures between radar viewpoints, thereby normalizing viewpoint-specific distortions while preserving motion-specific structures. Intuitively, this module performs a \emph{learned transformation of the radar “view”}, mapping the micro-Doppler representation of an activity as seen from radar~$m$ to how it would appear from radar~$n$. This allows downstream identity embeddings to be compared in a common, view-normalized representation space, enabling robust multi-radar re-ID.

\subsubsection{Pix2Pix-based View Transformation Network}
We adopt a conditional GAN architecture based on the \textit{Pix2Pix} framework~\cite{isola2017image}, which allows paired, supervised training of cross-view mappings. Unlike unpaired image translation methods such as CycleGAN~\cite{zhu2017unpaired}, Pix2Pix directly leverages aligned RD data collected from overlapping radar fields of view. The model is trained on a controlled indoor dataset where participants emulate typical stone-working activities such as cutting, chipping, grinding, and polishing while being simultaneously observed by two radars placed at different azimuthal separations. This setup provides paired RD spectrograms for the same motion instance viewed from multiple azimuth angles. The resulting training corpus spans a full $360^{\circ}$ coverage at $15^{\circ}$ increments, capturing the continuous transformation of micro-Doppler signatures with changing azimuth viewpoint. Let $\mathbf{H}_k^{(m)}$ denote the RD signature of user~$k$ observed by radar~$m$, and $\mathbf{H}_k^{(n)}$ the same activity observed by radar~$n$. The generator network $G_{m \rightarrow n}$ learns to transform $\mathbf{H}_k^{(m)}$ into the target view, $\hat{\mathbf{H}}_k^{(n)} = G_{m \rightarrow n}(\mathbf{H}_k^{(m)})$.
\subsubsection{Network Architecture}
The generator $G$ follows a \emph{U-Net} encoder–decoder architecture with skip connections, enabling the preservation of low-level spectral structures while learning global viewpoint transformations. The discriminator $D$ is implemented as a \emph{PatchGAN}~\cite{isola2017image}, which classifies local $70\times70$ patches rather than full images, encouraging high frequency consistency and reducing over smoothing in RD reconstructions. This architecture is well suited for RD spectrograms, where fine grained local textures encode micro-motion patterns.

\subsubsection{Loss Function}
We employ the Least-Squares GAN (LSGAN) objective~\cite{mao2017least} for improved training stability and gradient behavior compared to the original binary cross-entropy loss. The adversarial loss for the view transformation is given by:
\begin{equation}
    \mathcal{L}_{LSGAN}(G, D) = 
    \mathbb{E}_{\mathbf{H}^{(n)}}[(D(\mathbf{H}^{(n)}) - 1)^2] + 
    \mathbb{E}_{\mathbf{H}^{(m)}}[D(G(\mathbf{H}^{(m)}))^2].
\end{equation}

To enforce content consistency between the synthesized and target RD maps, we add an $L_1$ reconstruction term:
\begin{equation}
    \mathcal{L}_{L1}(G) = 
    \mathbb{E}_{\mathbf{H}^{(m)}, \mathbf{H}^{(n)}}[\|\mathbf{H}^{(n)} - G(\mathbf{H}^{(m)})\|_1].
\end{equation}

The total training objective is thus:
\begin{equation}
    \mathcal{L}_{total} = \mathcal{L}_{LSGAN}(G, D) + \lambda_{L1} \mathcal{L}_{L1}(G),
\end{equation}
where $\lambda_{L1}$ controls the trade-off between perceptual realism and structural fidelity. In our implementation, $\lambda_{L1}=100$ provided optimal convergence across all activity types.

\subsubsection{Effect of Azimuthal Dependence and Activity-Agnostic Learning}
The angular dependence of micro-Doppler patterns primarily manifests in the \textit{azimuth plane}, as most worker motions, such as arm swings, tool movements, and torso rotations, are planar and oriented parallel to the ground. In contrast, elevation-plane motion contributes minimally to the observed Doppler spread in typical stone-cutting and grinding postures. By learning transformations parameterized predominantly by azimuthal geometry, the network captures the underlying kinematic structure of motion rather than overfitting to activity semantics or environmental context. 

Consequently, the learned mapping becomes both \textit{activity-agnostic} and \textit{environment-agnostic}. Since the model is trained on diverse activity instances that share similar geometric dependencies across azimuth viewpoints, it generalizes to unseen or composite activities that follow comparable motion trajectories. The learned transformation thus reflects a fundamental property of radar scattering geometry rather than any specific action, allowing reliable view adaptation across different workplaces, tasks, and deployment geometries. We demonstrate this through field testing in different workshop environments, \textit{without retraining the pix-to-pix re-ID model} trained once in the lab setting (Section~\ref{sec:in-the-wild}).

\subsubsection{Effect of View Normalization}
Through this learned mapping, the generator produces view-normalized RD representations that preserve intrinsic motion signatures (e.g., chipping vs. grinding) while suppressing angle-induced Doppler distortions. As a result, embeddings derived from these normalized spectrograms become directly comparable across radar viewpoints, enabling robust correlation-based cross-radar re-ID (§\ref{sec:reid}) independent of activity type or environment.

\subsection{User Association and Re-Identification}\label{sec:reid}

To achieve consistent user identities across radars, we perform a two-stage association process combining pairwise matching and identity consolidation.  
For each frame, we compute the normalized correlation between activity signatures from radar $m$ and $n$. For user $a$ at radar $m$ and user $b$ at radar $n$, the correlation coeff. is:
\begin{equation}
    \rho(a_m, b_n) = \frac{\langle \hat{\mathbf{H}}_a^{(n)}, \mathbf{H}_b^{(n)}\rangle}{\|\hat{\mathbf{H}}_a^{(n)}\| \; \|\mathbf{H}_b^{(n)}\|},
\end{equation}
where $\hat{\mathbf{H}}_a^{(n)}$ is the view-adapted signature of user $a$ in radar $n$'s perspective. Pairs with $\rho(a_m, b_n) \geq \tau$ (empirically, $\tau=0.6$) are considered candidate matches.  

We construct a correlation matrix for all detections between the two radars, convert it to a cost matrix by subtracting from its maximum, and solve a one-to-one assignment using the Hungarian algorithm. 
Formally, for radar pair $(m,n)$ with $N_m$ and $N_n$ detections respectively, we define a similarity matrix $\mathbf{R} \in \mathbb{R}^{N_m \times N_n}$ where each entry corresponds to $\rho(a_m,b_n)$. Correlations below the threshold $\tau$ are suppressed prior to assignment. The matrix is converted into a cost matrix $\mathbf{C}$ such that $C_{a,b} = \max(\mathbf{R}) - R_{a,b}$, ensuring that higher similarity results in lower assignment cost. Entries corresponding to suppressed correlations are assigned a large penalty to prevent forced matches.
 We apply a correlation threshold and optionally enforce a mutual-best check, retaining only the strongest bidirectional correspondences. 
The Hungarian algorithm is applied to cost matrix $C$ to derive a globally optimal one-to-one assignment that minimizes the total cost objective. Subsequently, assignments with a correlation coefficient below the threshold $\tau$ are pruned. In high-density scenarios, an optional mutual-best constraint is enforced to mitigate ambiguous associations.
 The surviving pairs constitute unique matches for that frame. Across multiple frames and/or additional radars, we aggregate all accepted pairwise matches into an undirected graph, where nodes represent local detection IDs (e.g., ``U11'', ``U21'') and edges represent confirmed matches. Connected components of this graph define transitive equivalence classes of detections belonging to the same user. Each component is assigned a deterministic global user ID (P1, P2, ...), and all member detections are recorded. 
To ensure identity persistence over time, this association graph is maintained across consecutive frames. Nodes are time-stamped, and short-term missed detections are tolerated within a predefined temporal window. If a user reappears within this window and within a spatial proximity threshold of its last known global position, the previous identity is reactivated to prevent identity fragmentation.

This pipeline guarantees strict one-to-one matches at the pairwise stage, robust multi-radar/multi-frame clustering via transitive closure, and stable global IDs suitable for downstream tracking. In our case, the setup involves three radars participating in the pairwise association process. Sequentially, radar 1's transformed signatures are correlated with radars 2 and 3, followed by radar 2 correlated with radar 3. The resulting global associations are stored in a matrix $\mathbf{A}$, where entry $\mathbf{A}_{m,k} = g$ indicates that user $k$ at radar $m$ is assigned to global ID $g$.

\subsection{Use-case Design: Global User Positioning and PM Exposure Estimation}

Using the globally consistent user identities obtained from cross-radar association, we project all radar-localized coordinates into a unified global coordinate frame. We select one radar node, denoted as radar $r_0$, as the reference and define all transformations relative to it. Let $\mathbf{T}_m$ represent the rigid-body transformation (rotation and translation) from radar~$m$’s local coordinates to the reference radar’s coordinate system. The global position of user~$k$ observed by radar~$m$ is computed as $\mathbf{u}_k^{global} = \mathbf{T}_m \cdot \mathbf{u}_k$.

Once all users’ trajectories are expressed in this common frame, we estimate their personalized exposure to PM. A sparse network of PM sensors (\textit{DALTON}~\cite{karmakar2024exploring}) distributed across the workspace provides localized pollutant concentration readings $\{PM_j(\mathbf{s}_j)\}$ at sensor locations $\mathbf{s}_j$. Conceptually, the continuous PM concentration field can be modeled via inverse distance weighting (IDW):
\begin{equation}
    \hat{PM}(\mathbf{x}) = 
    \frac{\sum_j w_j(\mathbf{x}) PM_j}{\sum_j w_j(\mathbf{x})},
    \qquad 
    w_j(\mathbf{x}) = \frac{1}{\|\mathbf{x}-\mathbf{s}_j\|^p},
\end{equation}
where $p$ controls the decay rate with distance and $\hat{PM}(\mathbf{x})$ denotes the interpolated pollution value at position $\mathbf{x}$.

In practice, since the number of PM sensors is limited, we approximate $\hat{PM}(\mathbf{x})$ using a discrete spatial heatmap representation rather than a fully continuous interpolation, as followed in several existing literature~\cite{liu2018third,wu2020sharing}. Each PM sensor is associated with a spatial zone in the workspace corresponding to its physical placement, and the measured pollution levels are projected onto these zones. Zone-level pollution values are then smoothed to form a coarse pollution field, providing an interpretable spatial distribution of PM concentration over the monitored area. This implementation captures the relative concentration gradients necessary for per-worker exposure estimation while avoiding instability from sparse-sensor interpolation.
For each user trajectory $\{\mathbf{u}_k^{global}(t)\}$, the personalized exposure is estimated as:
\begin{equation}
    E_k = \frac{1}{T}\int_0^T \hat{PM}\big(\mathbf{u}_k^{global}(t)\big) dt,
\end{equation}
where $T$ is the observation duration. The resulting $E_k$ values provide individualized, time-averaged exposure estimates by combining radar-tracked trajectories in the global frame with the discretized PM field. 
Radar localization and re-identification operate at 10\,Hz, while PM sensors sample at 1\,Hz. For synchronization all streams are timestamp-aligned, and exposure is computed over a 5-second aggregation window. Within each window, we compute the median re-ID outputs and corresponding PM readings before estimating $E_k$. This temporal smoothing reduces the impact of transient mismatches and sensor noise.

\section{Implementation}\label{impl}
The implementation of \ourmethod integrates a multi-radar network, micro-controller-based sensor interfaces, and a software pipeline consisting of data acquisition, pre-processing, and view adaptation. 

\begin{figure*}[t]
    \centering
    \subfigure[Lab setup with Vicon cameras]{
    \includegraphics[width=0.25\textwidth]{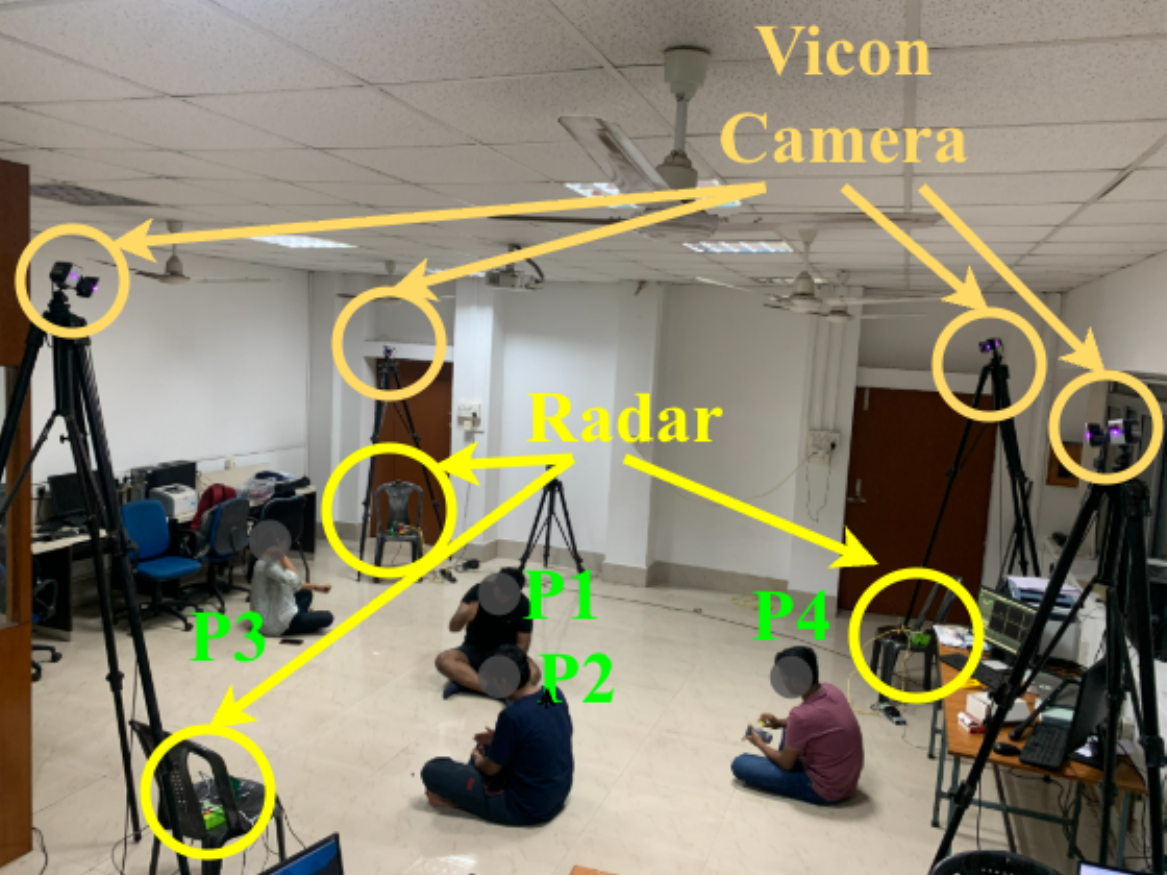}
    \label{fig:impl_vicon_lab}
    }\hfil
    \subfigure[Marble processing factory (outdoor)]{
    \includegraphics[width=0.24\textwidth]{figures/StoneWorkerBD-marblefactory.drawio.pdf}
    \label{fig:impl_marble_outdoor}
    }\hfil
    \subfigure[Stone-cutting factory]{
    \includegraphics[width=0.165\textwidth]{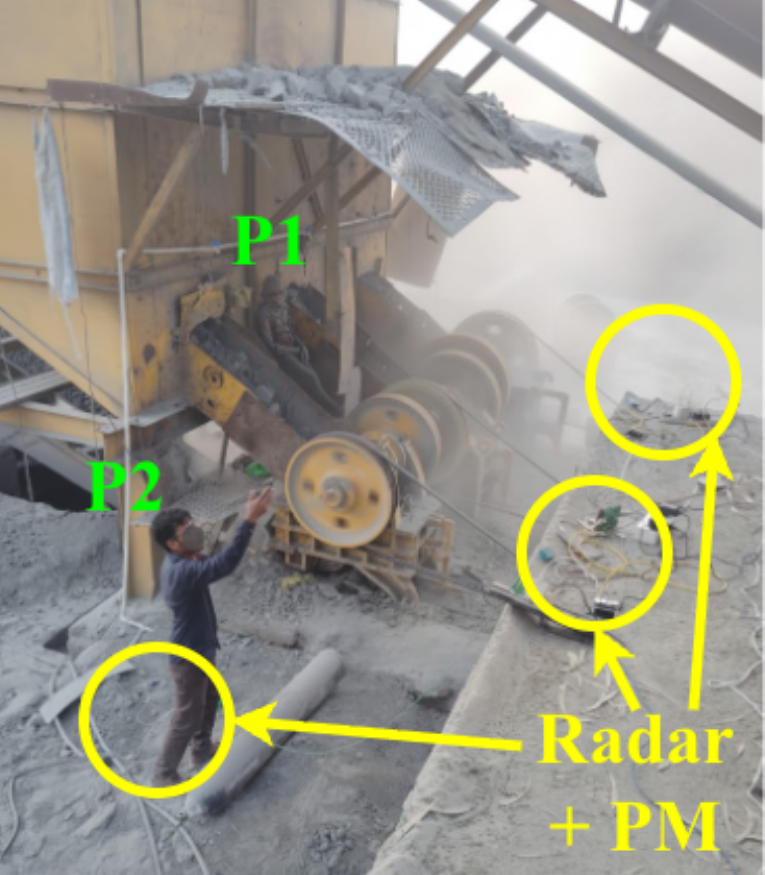}
    \label{fig:impl_khadan}
    }\hfil
    \subfigure[Indoor construction workspace]{
    \includegraphics[width=0.21\textwidth]{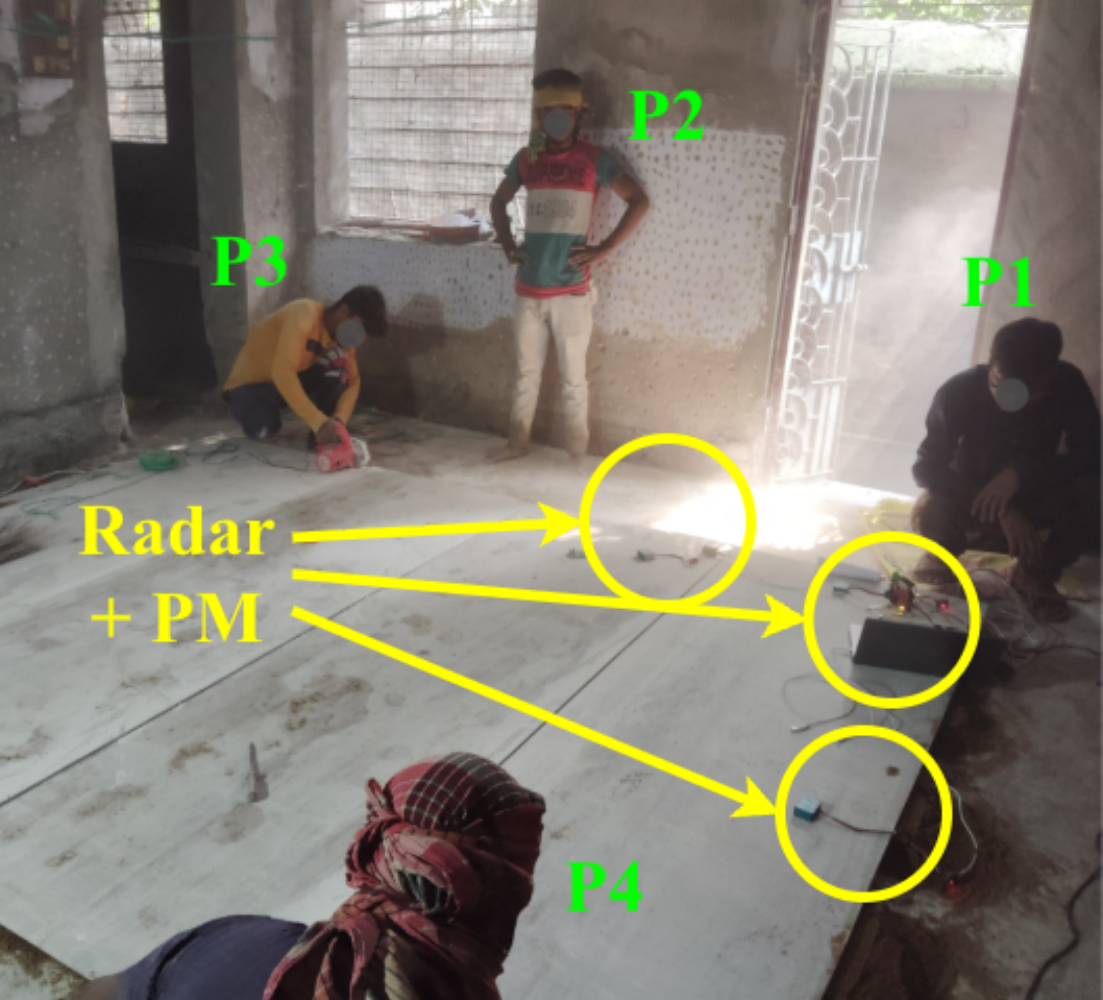}
    \label{fig:impl_indoor}
    }
    \caption{Multiple deployment scenarios of \ourmethod.} 
    \label{fig:clear_deployments}\vspace{-0.4cm}
\end{figure*}

\subsection{Hardware Setup}
Each sensing node is built with Texas Instruments IWR1843BOOST mmWave radar~\cite{iwr1843boost}, which provides FMCW based range and Doppler signatures. The raw ADC data were captured using DCA1000EVM~\cite{dca1000evm} data captured card and streamed over Gigabit Ethernet to a local embedded processor. Alongside the radar, we deployed an ESP32 NodeMCU microcontroller for handling serial and digital interfaces with the pollution sensors. We employed an NVIDIA Jetson Nano at each node as the central compute unit, receiving radar streams over ethernet and sensor data over Wi-Fi. Figure~\ref{fig:node} shows one deployed radar–sensor node with its ESP32 and Jetson Nano stack. The radars were configured with TI mmWave Studio software. We used a chirp profile sweeping 4~GHz bandwidth (77-81~GHz) with a chirp duration of $72$ $\mu$s and a sampling rate of $4400$~Msps. Each frame contained $182$ chirps at a frame rate of 10~Hz with approximately $4$~cm range resolution and $182$ Doppler resolution to capture micro-motion signatures.

\subsection{Software Framework}
 
The collected raw ADC data from the Ethernet streams are first converted to RD heatmaps and pointclouds using custom Python scripts. The ESP32 NodeMCU is programmed using the Arduino framework, enabling serial data collection from PM and gas sensors, packetization, and transmission over Wi-Fi. On the receiving end, the Jetson Nano runs a lightweight server that synchronizes incoming Wi-Fi sensor streams with Ethernet radar streams. TI mmWave Studio is used for radar configuration. All collected data are stored locally on the Nano and mirrored to a central server for offline processing and model training. The view-adaptation pipeline is trained on a Tesla~V100 GPU (two 16\,GB GPUs) in a rack server with 256\,GB RAM and an \texttt{Intel Xeon(R) Gold 6152} CPU with 88 cores. The train–test split used is 80:20.
\subsection{Deployment Scenarios}

We evaluated \ourmethod in two distinct phases: (1) laboratory experiments with precise ground truth for validation of view-adaptation and re-ID pipeline, and (2) real-world field deployments.

\subsubsection{Laboratory Validation Setup}
We recruited $15$ participants and instructed them to emulate typical stone-cutting and polishing motions, including chipping, grinding, and polishing gestures. The ground-truth spatial coordinates of each participant were captured using a Vicon motion capture system with five infrared cameras operating at 100~Hz, providing sub-millimeter accuracy. This setup enabled us to evaluate the end-to-end accuracy of the radar-based localization and the consistency of our view-adaptation and re-ID pipeline against high-precision user positioning. Furthermore, to train the view-adaptation network described in Section~\ref{sec:view_adaptation}, we explicitly collected paired radar data from two radar nodes positioned at controlled azimuthal separations. One radar was fixed at $0^{\circ}$ reference orientation, while the second radar was sequentially placed at $15^{\circ}$, $30^{\circ}$, $45^{\circ}$, and so on up to $360^{\circ}$. At each angular configuration, participants performed the same set of activities to capture paired RD representations of identical motions from multiple viewpoints. Each activity session lasted approximately 3–5 minutes per angular configuration at 10 FPS, for an effective cumulative recording duration of approximately 6 hours and yielding over 126,000 RD frames used for supervised training. This systematic multi-angle data collection enabled supervised training of the Pix2Pix-based view transformation model, enabling it to learn azimuthal dependencies and translate robustly across multiple radar perspectives. Following training in a lab-scale setup, we evaluated the pretrained model in real-world deployment scenarios, as discussed next.

\subsubsection{Real-World Deployments}\label{sec:deployments}
To assess robustness under actual working conditions, we deployed \ourmethod across three different stone-processing environments, as follows.  

\begin{sitemize}
    \item \textbf{Site~1: Marble Processing Factory (Outdoor).} 
    This large open-air marble factory hosts multiple work zones where $5$ stone workers simultaneously perform cutting, chipping, grinding, and polishing operations on marble blocks. The workspace features continuous dust emission and overlapping worker trajectories, posing challenges for radar visibility. 

    \item \textbf{Site~2: Stone-Cutting Factory.} 
    This site represents a mechanized stone-cutting factory where dust generation originates primarily from high-speed stone-cutting machines. Despite only two workers, continuous machinery and heavy dust made it ideal for testing robustness of the re-ID pipeline.

    \item \textbf{Site~3: Indoor Construction Workspace.}
    The third deployment took place in a partially enclosed indoor construction area where marble slabs were being cut and installed, with four workers present. Unlike the outdoor sites, this environment had strong multipath and limited fields of view. 
\end{sitemize}

Across all sites, three mmWave radar nodes were deployed in partially overlapping configurations, synchronized through Ethernet, and co-located with PM sensors. The system continuously captured multi-radar RD data streams and environmental PM readings. The overlapping radar placement ensures users remain observable despite \textit{occlusions} or \textit{FoV limits}, as at least one other node can maintain visibility. For example, P4 in \figurename~\ref{fig:impl_indoor} is occluded in Radar 1's FoV but remains visible to the other nodes.

Real-world data, totaling 20 hours and over 265,000 RD frames, were recorded across three sites to capture varying environmental conditions. The view-adaptation model, trained on laboratory data (80:20 split), was evaluated in these unseen field environments without fine-tuning to test generalization. To ensure rigorous re-identification assessment, a participant-independent protocol was used, keeping training and testing identities mutually exclusive.
\section{Lab-Scale Evaluation}
We first evaluate the overall performance of \ourmethod\ in the lab setup, followed by component-wise analysis to quantify their individual accuracy and robustness against baseline methods.


\subsection{Overall Performance}
\label{sec:overall_performance}
To quantify the end-to-end performance of \ourmethod{}, we compare it against two baseline re-ID strategies: 
(i) a \textit{distance-based} approach that associates detections across radars using nearest-neighbor matching in global coordinates, and 
(ii) a \textit{correlation-based} approach that uses micro-Doppler activity correlation between radar nodes without any view normalization. Notably, to the best of our knowledge, our method is the first to be able to identify an object through multiple radars at the same time. Thus, we compare these two classical methods for radar point-cloud to human-subject association with our approach under multi-radar setup. 

\begin{figure}[!t]
    \centering
    \subfigure[Re-ID F1-Score across \# of users]{
        \includegraphics[width=0.21\textwidth]{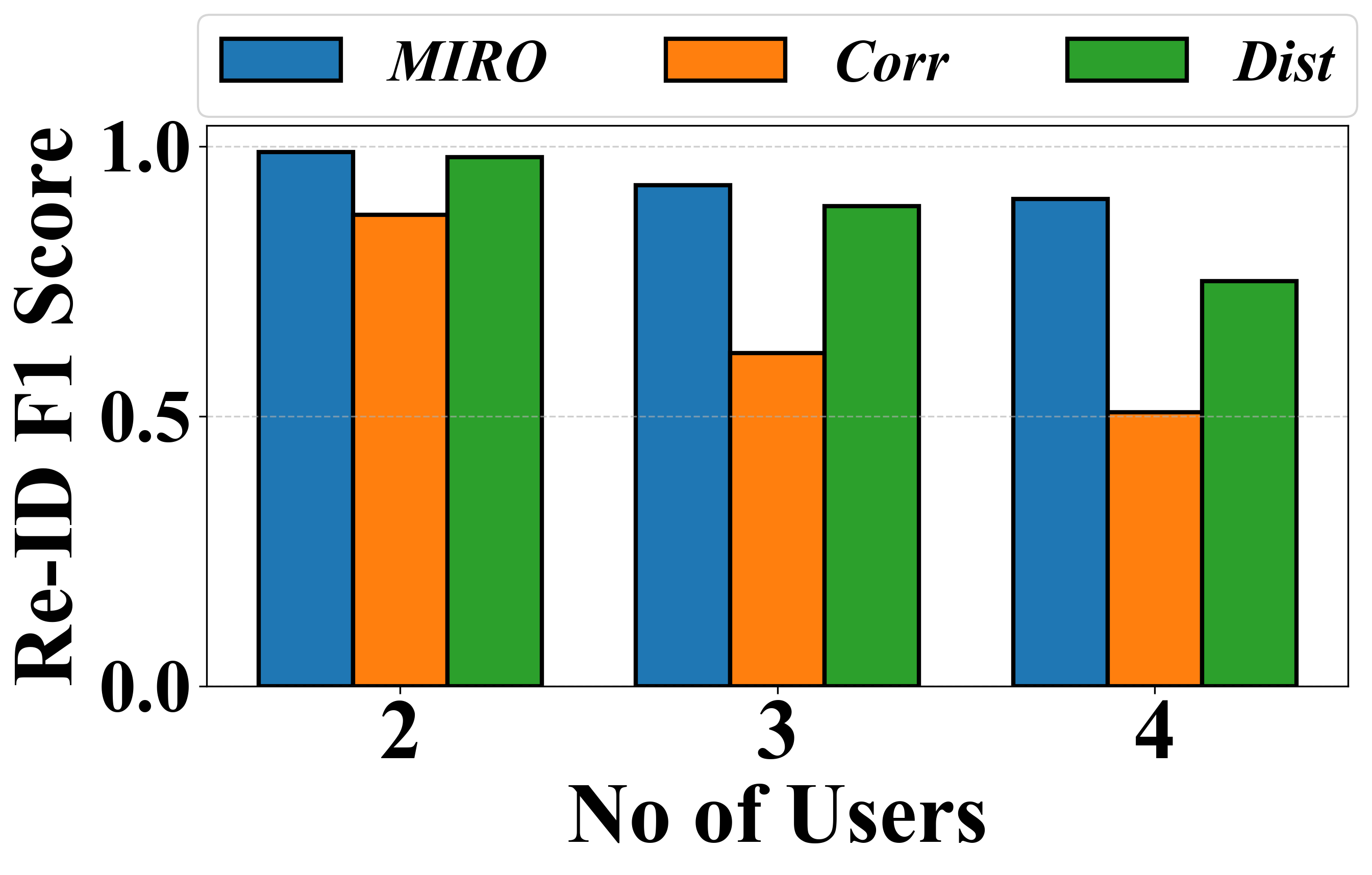}
        \label{fig:overall}
    }\hfil
    \subfigure[Confusion matrix of re-ID]{\includegraphics[width=0.18\textwidth]{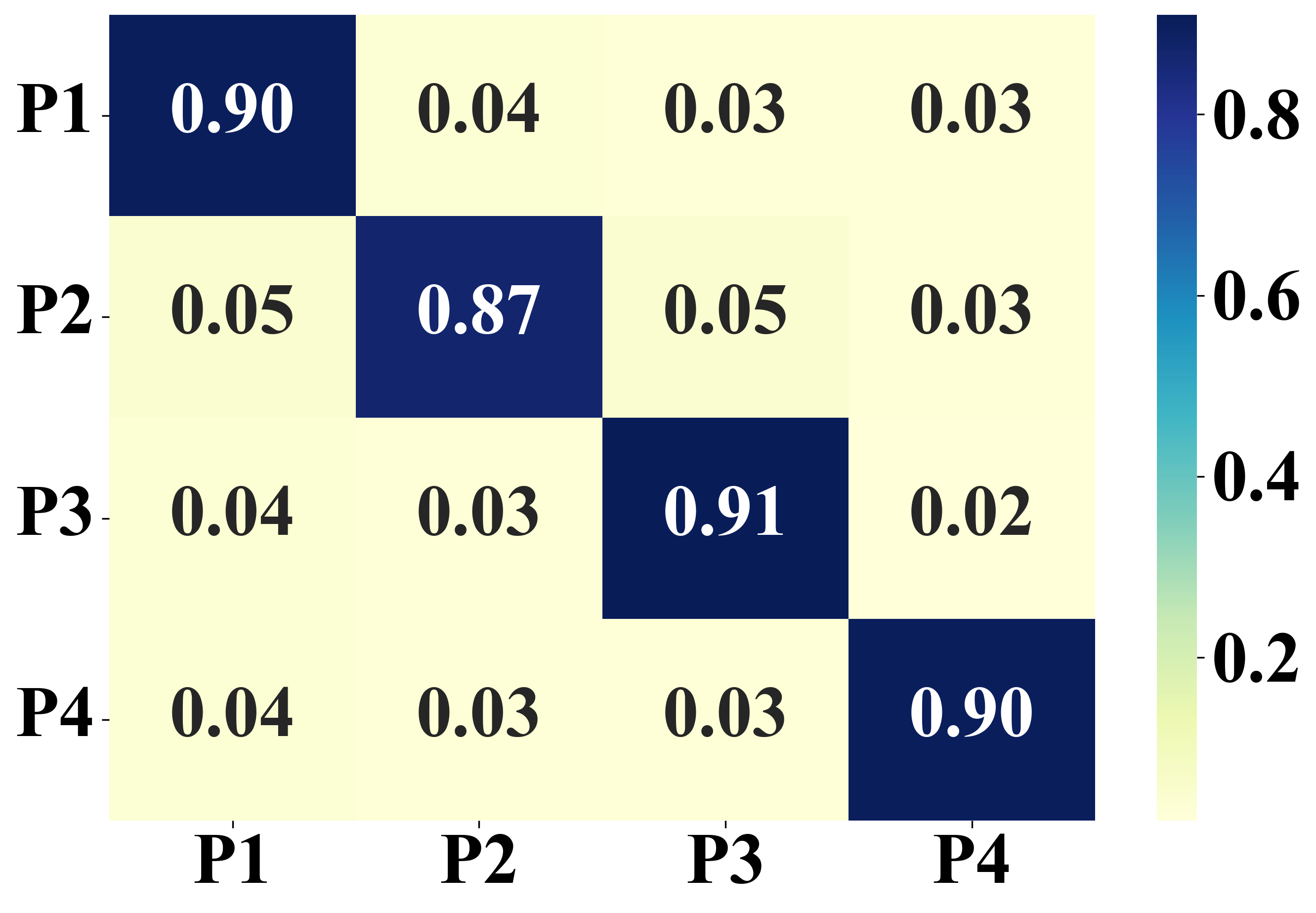}\label{fig:personReIDConfusionMatrix}}
    \caption{Overall re-ID performance.}
    \vspace{-0.45cm}
\end{figure}

As shown in \figurename~\ref{fig:overall}, the \textit{distance-based} baseline suffers with increasing user count (from $0.91$ for two to $0.73$ for four users), as it relies on one-to-one Hungarian matching in global coordinates and becomes unreliable under higher densities due to localization noise and spatial overlap. While the \textit{correlation-based} baseline fails under dense settings as unaligned viewpoints distort temporal and angular features. Lacking view adaptation, pure correlation leads to frequent identity confusion. In contrast, \ourmethod{} sustains near-constant F1 scores across densities by employing a \textit{Pix2Pix-LSGAN–based view adaptation} that learns azimuth-aware geometric transformations, producing view-consistent, activity-invariant RD representations.  Consequently, \ourmethod{} achieves superior scalability and robustness in dense multi-radar deployments, as evidenced by the per-user consistency in \figurename~\ref{fig:personReIDConfusionMatrix}. 
\ourmethod targets small-scale industrial work zones with a modest number of concurrent workers ($\approx 3-6$). While Fig.~\ref{fig:overall} evaluates up to four users, performance may degrade at higher densities due to radar resolution limits and overlapping range–Doppler signatures.

\subsection{Microbenchmark Evaluation}
We evaluated the performance of clustering, view adaptation, and cross-radar re-ID components of \ourmethod{}.

\begin{figure}[!ht]
    \centering
    \subfigure[Cluster Count Accuracy]{
    \includegraphics[width=0.17\textwidth]{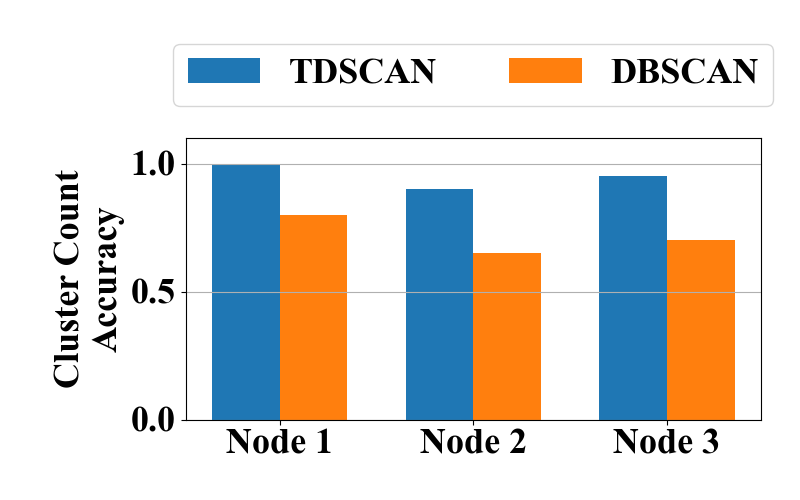}\label{fig:clustertdscan}
    }\hfil
    \subfigure[MAE per Radar Node]{
    \includegraphics[width=0.14\textwidth]{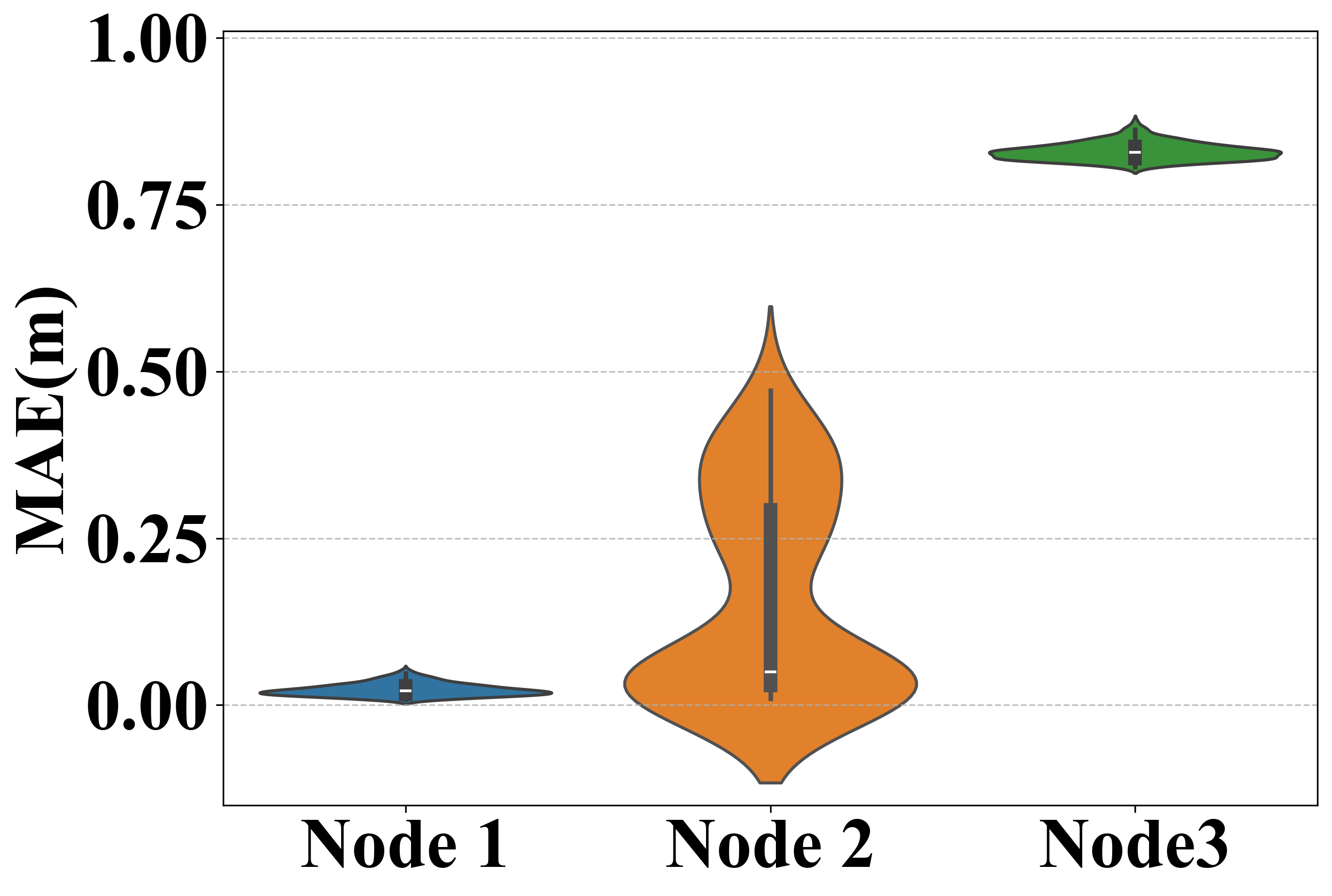}
    \label{fig:mae_node}
    }\hfil
    \subfigure[MAE per Subject]{
        \includegraphics[width=0.13\textwidth]{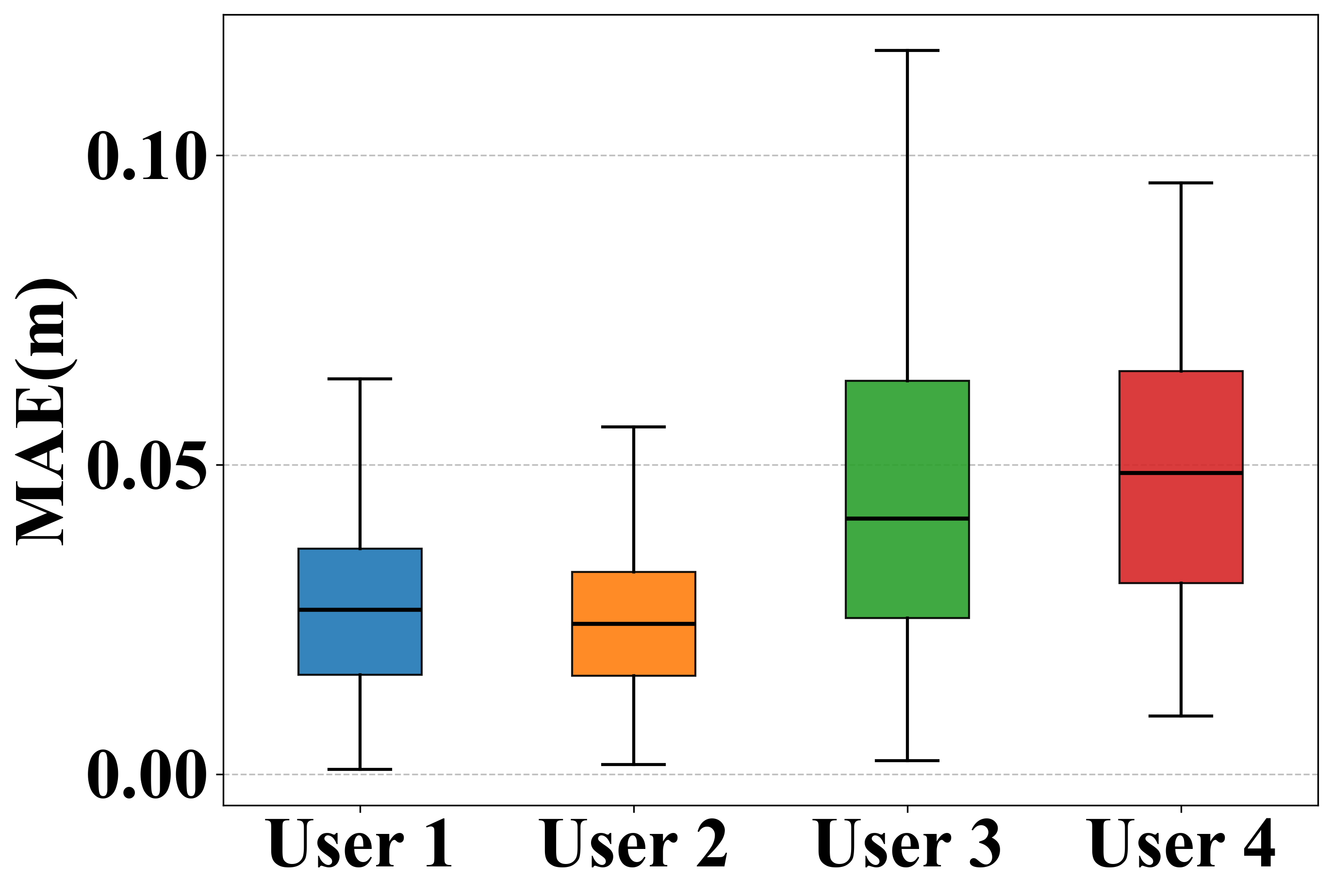}
        \label{fig:mae_user}
    }
    \caption{Comparison of radar-based position estimates against Vicon measurements (for the setup of \figurename~\ref{fig:impl_vicon_lab}).}\vspace{-0.5cm}
\end{figure}

\subsubsection{Performance of Clustering}
We first examine the accuracy of clustering pointclouds into user instances. Clustering performance is measured as $Acc = 1 - \frac{|C_{detected}-C_{actual}|}{C_{actual}}$, where $C_{detected}$ is the number of clusters formed and $C_{actual}$ the ground-truth count of workers. As shown in Fig.~\ref{fig:clustertdscan}, the proposed TDSCAN method consistently achieves high accuracy compared to standard DBSCAN based approach. As a result of TDSCAN's design, which employs Doppler-based thresholding to suppress static clutter, and then stacks pointcloud frames in a timely manner, it creates a cleaner and more coherent representation of each worker. Beyond cluster detection, we assess how accurately TDSCAN estimates the spatial positions of the clusters. The mean absolute error (MAE) between each cluster centroid and its corresponding Vicon-tracked ground truth is summarized in Fig.~\ref{fig:mae_node} and Fig.~\ref{fig:mae_user}.

Node-wise, we observe that Node3 can see four cluster IDs, resulting in a higher accumulated error compared to the other nodes. Node2 and Node1 only capture two cluster IDs each, producing similar but comparatively lower accumulated errors. Across all users, the average MAE remains below $10$~cm, with slightly higher errors observed for users positioned near the radar's field-of-view (FoV) boundaries due to sparser point returns. Increased user radar distance also contributes to higher errors, as it raises the likelihood of intermediate interference from other users' activities entering the FoV and degrading the pointcloud density of the farthest user. In our setup, \textit{user~4} was located at the largest distance from the radar ($6$~m), whereas \textit{users~1-3} were positioned within $3$~m.

For a single radar view, lateral position estimates rely on azimuthal angle measurements; thus, even a small angular error $\Delta\theta$ translates into a lateral displacement of approximately $R\Delta\theta$, which increases linearly with range. Moreover, radar path loss at longer distances reduces return power, lowers SNR, and increases CFAR misses, yielding sparser and noisier point sets. Under such conditions, DBSCAN becomes more fragile: clusters may fragment if \texttt{min\_samples} is set too high, or may merge incorrectly if \texttt{eps} is enlarged to preserve cluster continuity. 

\begin{figure}[!ht]
    \centering
    \subfigure[Neighborhood radius $\epsilon$]{
    \includegraphics[width=0.2\textwidth]{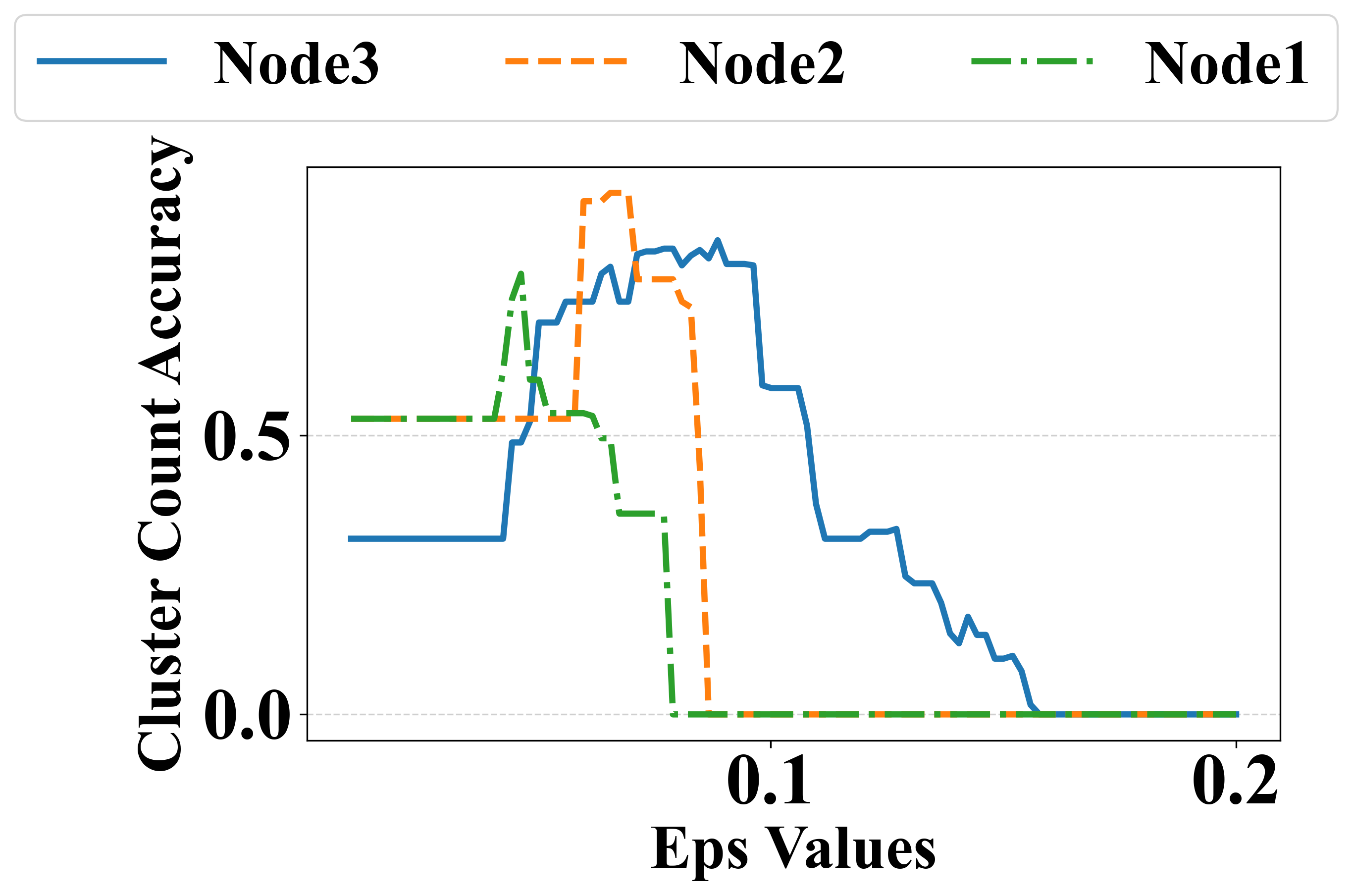}
    \label{fig:acc_eps}
    }\hfil
    \subfigure[Minimum number of points]{
        \includegraphics[width=0.2\textwidth]{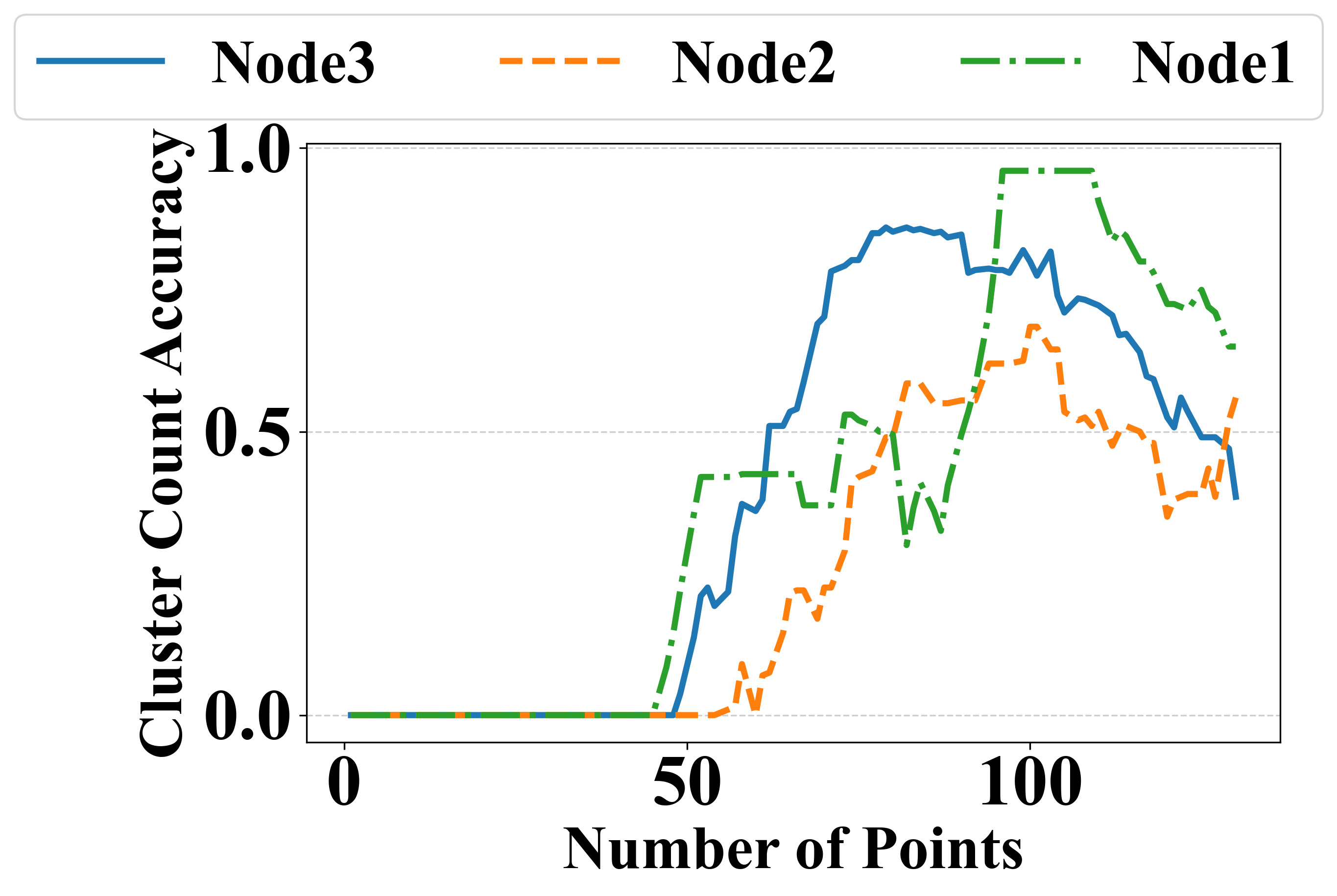}
        \label{fig:acc_points}
    }
    \caption{Clustering count accuracy of TDSCAN parameters.}\label{fig:clustering_param}\vspace{-0.5cm}
\end{figure}

After characterizing the performance of TDSCAN, we next tune the algorithm's parameters by varying both the neighborhood radius $\epsilon$ and the minimum number of points (\texttt{minPts}) (see  \figurename~\ref{fig:clustering_param}). We observe a clear optimum at $\epsilon = .75$ m and minPts = $100$ in \figurename~\ref{fig:acc_eps} and \figurename~\ref{fig:acc_points} respectively, achieving an average cluster count accuracy of $94\%$. These parameters correspond well to the physical spacing between individuals in the worksite. 

In this environment, each worker operates in close proximity to their stone block and tools, forming a compact point cloud roughly matching the size of a human body (around 0.5-1 m in spatial extent). If $\epsilon$ is too small, the algorithm splits a single worker into multiple fragments (for example, separating the body and nearby tools). Conversely, if $\epsilon$ is too large, neighboring workers may be merged into one cluster. The optimal value of $\epsilon = 0.75$m effectively captures worker bodies and instruments, resulting in clusters that tightly enclose each individual's workspace.


\begin{table}[]
\centering
\scriptsize
\caption{Performance of different view-adaptation methods.}\label{tab:viewadapt}
\begin{tabular}{|c|c|c|c|}
\hline
\textbf{Model}           & \textbf{L1\_mean} & \textbf{SSIM\_mean} & \textbf{PSNR\_mean} \\ \hline
CycleGAN   & 12.8522  & 0.4967     & 19.48      \\ \hline
Pix2Pix WGAN-GP & 19.4237  & 0.2654     & 18.27      \\ \hline
\textbf{Pix2Pix LSGAN} & \textbf{6.1296}   & \textbf{0.6988}     & \textbf{28.02}      \\ \hline
\end{tabular}
\vspace{-0.4cm}
\end{table}

\subsubsection{Performance of View-Adaptation}

\begin{figure}
    \centering
    \includegraphics[width=0.46\textwidth]{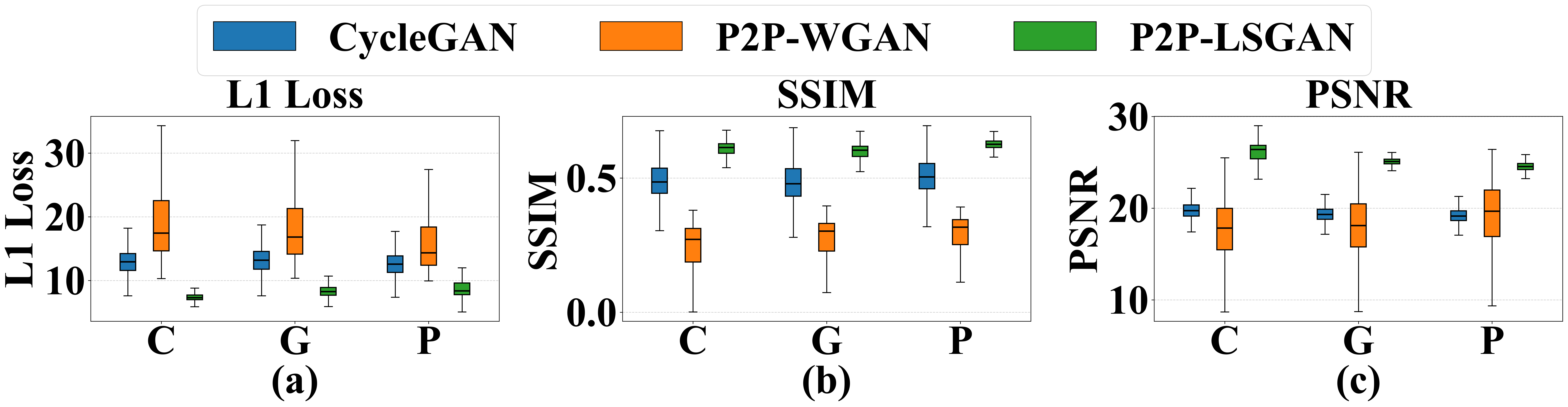}
    \caption{View adaptation performance: (a) L1-Loss, (b) SSIM, (c) PSNR (C: Chipping, G: Grinding, P: Polishing).}
    \label{fig:matric}
\end{figure}

We compared three image-to-image translation frameworks for cross-view radar representation learning: \textit{Pix2Pix with Least Squares GAN (Pix2Pix-LSGAN)}~\cite{isola2017image}, \textit{Pix2Pix with Wasserstein GAN and Gradient Penalty (Pix2Pix-WGAN-GP)}~\cite{gulrajani2017improved}, and \textit{CycleGAN}~\cite{zhu2017unpaired}. Each model was trained to translate RD heatmaps between radar viewpoints, enabling consistent micro-Doppler interpretation across different azimuth angles. To quantify view-adaptation quality, we evaluate both pixel-level reconstruction fidelity and structural preservation using $L_1$ loss, Structural Similarity Index (SSIM), and Peak Signal-to-Noise Ratio (PSNR). As shown in \figurename~\ref{fig:matric}(a)--(c) and \tablename~\ref{tab:viewadapt}, the \textit{Pix2Pix-LSGAN} model achieves the best overall performance, with a mean $L_1$ loss of $6.12 \pm 0.12$, SSIM exceeding $0.69$, and PSNR above $28.0$~dB across all activities. 

The superior performance of \textit{Pix2Pix-LSGAN} can be attributed to its stable adversarial formulation and balanced optimization objective. The least-squares loss penalizes samples that deviate from real RD heatmaps more smoothly than the binary cross-entropy in standard GANs. In contrast, the WGAN-GP variant introduces strong gradient constraints that can suppress high-frequency doppler details, while CycleGAN, often struggles to maintain pixel-wise consistency across views. Consequently, \textit{Pix2Pix-LSGAN} produces sharper heatmaps with a better trade-off between adversarial realism and reconstruction accuracy across various radar views.


We further analyzed the per-activity adaptation behavior to interpret how motion patterns in each activity influence view-adaptation. As shown in \figurename~\ref{fig:matric} activities such as \textit{grinding} yield more consistent adaptation than \textit{polishing}, possibly as a result of their broader Doppler distributions. This indicates that GANs benefit from motion patterns with richer frequency content, as they provide stronger spatial-temporal cues for domain alignment. In contrast, \textit{Pix2Pix}, which is primarily designed for paired image-to-image translation, captures the underlying structural patterns of RD signatures without a need for spatial correspondence. This is particularly important for RD data, where the global energy distribution over velocity and range conveys more discriminative information than exact pixel-level alignment. 
Overall, \textit{Pix2Pix-LSGAN} demonstrates superior performance compared to both \textit{CycleGAN} and \textit{Pix2Pix-WGAN-GP}, which suffer from weaker structural consistency or over-regularization. By jointly enforcing adversarial realism and pixel-level reconstruction, \textit{Pix2Pix-LSGAN} learns robust azimuth-invariant mappings that help for reliable cross-radar re-ID.

\begin{figure}
    \centering
    \subfigure[F1-score for different methods]{
        \includegraphics[width=0.19\textwidth]{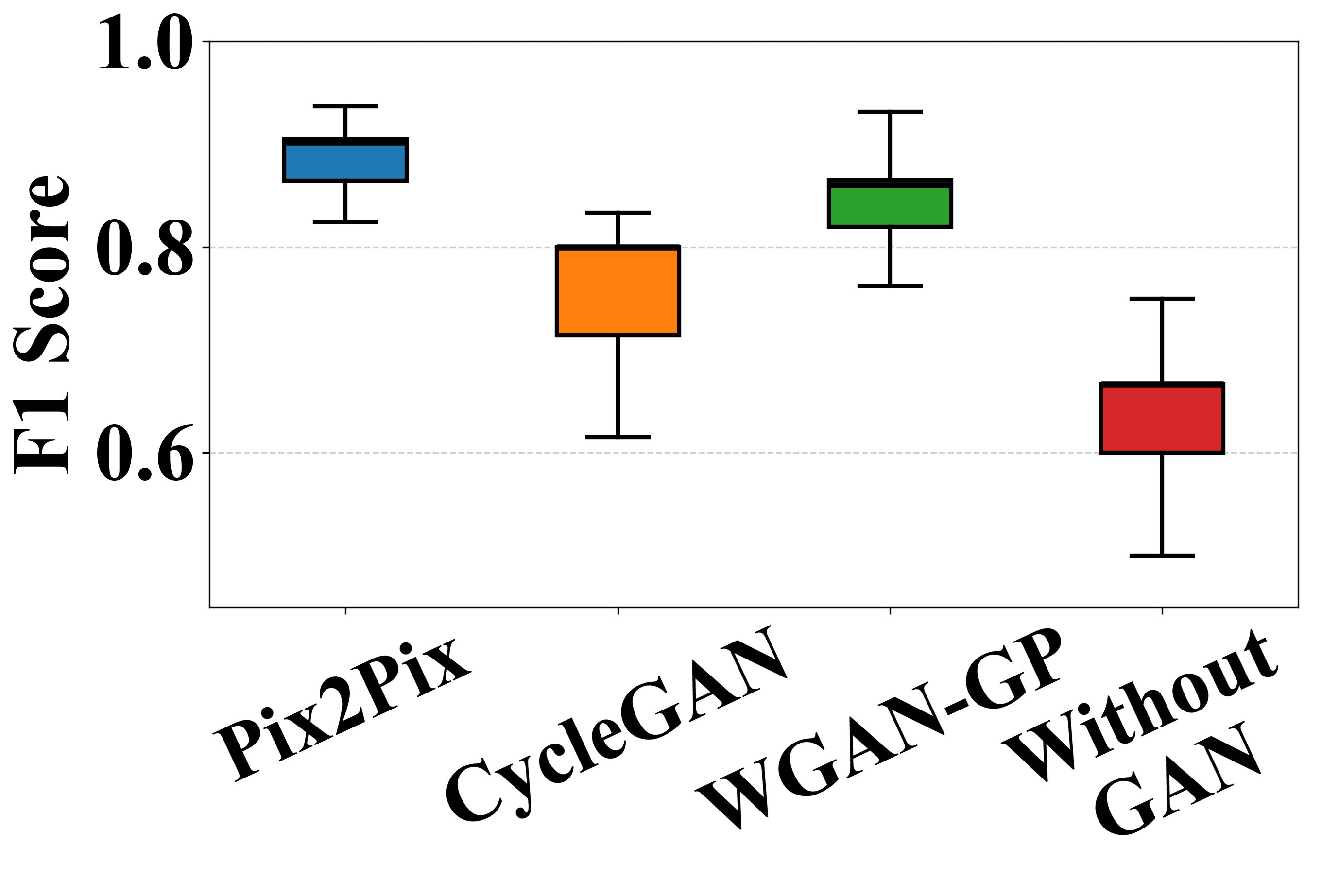}
        \label{fig:reid-performance}
    }\hfil
    \subfigure[Pipeline latency breakdown]{\includegraphics[width=0.19\textwidth]{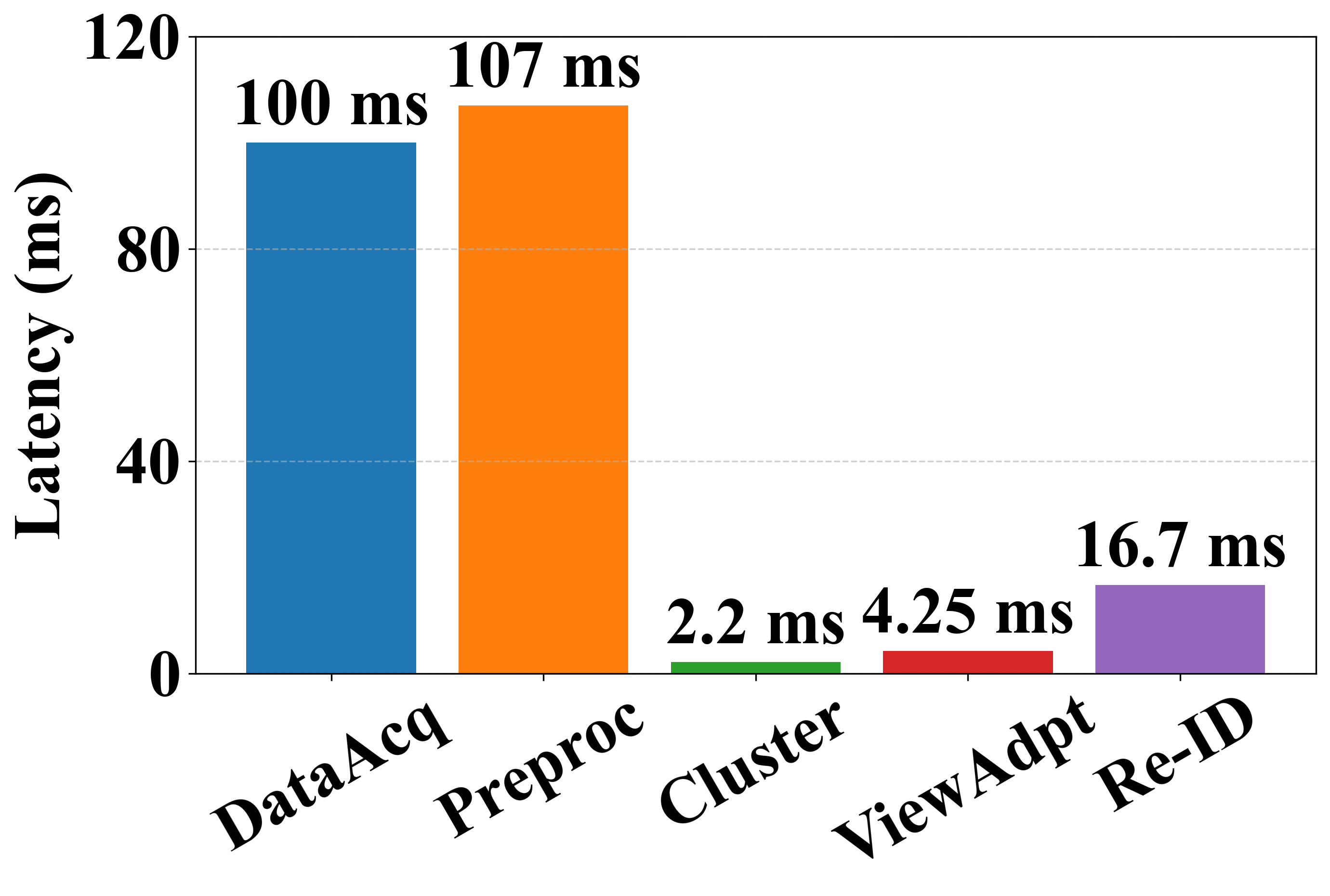}\label{fig:latency_donut}    
    }
    \caption{
        Performance evaluation of \ourmethod.}
    \vspace{-0.5cm}
\end{figure}



\subsubsection{Performance of Re-Identification}
We further assess the impact of different view-adaptation on multi-radar identity association. As summarized in \figurename~\ref{fig:reid-performance}, the proposed fusion model employing \textit{Pix2Pix-LSGAN} achieves a median re-ID F1-score of $90.4\%$ across three radar nodes, outperforming both \textit{Pix2Pix-WGAN-GP} ($86.8\%$) and \textit{CycleGAN} ($80\%$). As a result, LSGAN-based view normalization enhances cross-radar matching by producing geometrically consistent and activity-invariant representations. 

When GAN-based view adaptation is removed, performance drops by approximately 20\%, highlighting the crucial role of viewpoint alignment in sustaining identity continuity. Overall, these results confirm that \ourmethod{}, when integrated with \textit{Pix2Pix-LSGAN}–driven view adaptation, delivers the most stable and accurate re-ID across spatially distributed radar views.

\subsubsection{Response Time Analysis}\label{sec:latency}
\figurename~\ref{fig:latency_donut} presents the end-to-end latency breakdown of the proposed system across its major processing stages. The overall response time is primarily dominated by the Data Acquisition (DataAcq) and Preprocessing (Preproc) stages, which account for 100 ms and 107 ms, respectively. The data acquisition latency is mainly constrained by hardware, as our mmWave radar operates at 10 Hz. In contrast, the computational modules exhibit significantly lower latency. The Clustering module requires only 2.2 ms, while the View Adaptation (ViewAdapt) module completes within 4.25 ms. The final re-ID stage adds 16.7 ms. The complete system has a latency of ~230 ms per prediction (excluding a 1-s initial buffer bootstrapping). Since stone-worker activity changes over minutes to hours, this latency is easily acceptable.


\section{In-the-wild Evaluation}
\label{sec:in-the-wild}
To assess the generalization of \ourmethod{} beyond the controlled lab setup, we deployed the pretrained view-adaptation model across unseen activities and three unseen environments (\S\ref{sec:deployments}). We focus primarily on view-adaptation performance, as Vicon-based reference collection was infeasible in field conditions.

\subsection{View-Adaptation under Unseen Activities}
To further evaluate the activity invariance of the proposed view-adaptation model, we tested it on three previously \textit{unseen actions} -- \textit{clapping}, \textit{waving}, and \textit{nodding}, captured from a single radar viewpoint. These activities were not part of the training dataset, which originally included only industrial gestures emulated in the lab environment (\textit{chipping}, \textit{grinding}, and \textit{polishing}). 

\begin{figure}[!ht]
    \centering
    \subfigure[Unseen activities]{\includegraphics[width=0.19\textwidth]{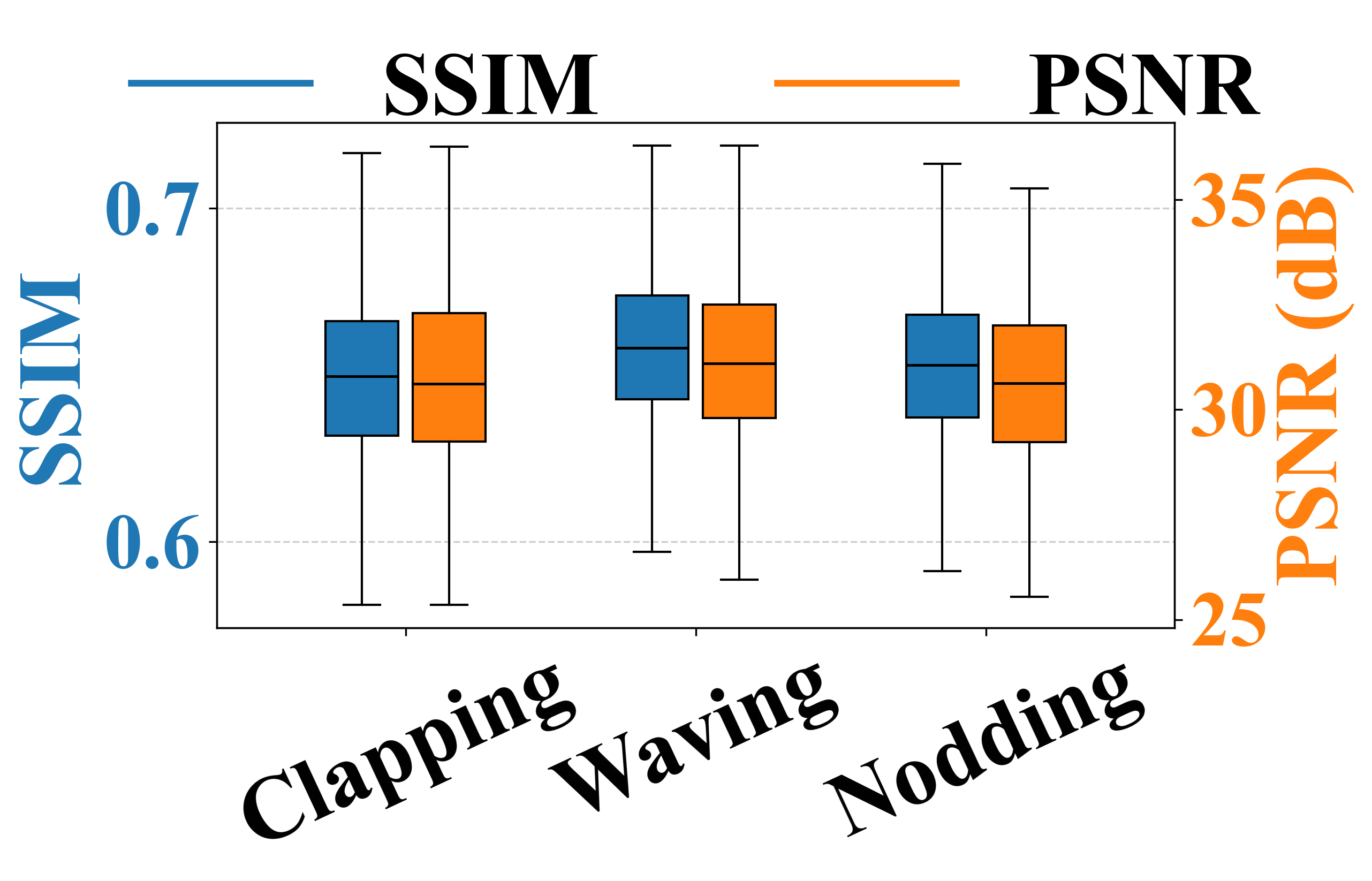}    \label{fig:unseenActivityBox}
    }\hfil
    \subfigure[Unseen deployments]{
    \includegraphics[width=0.19\textwidth]{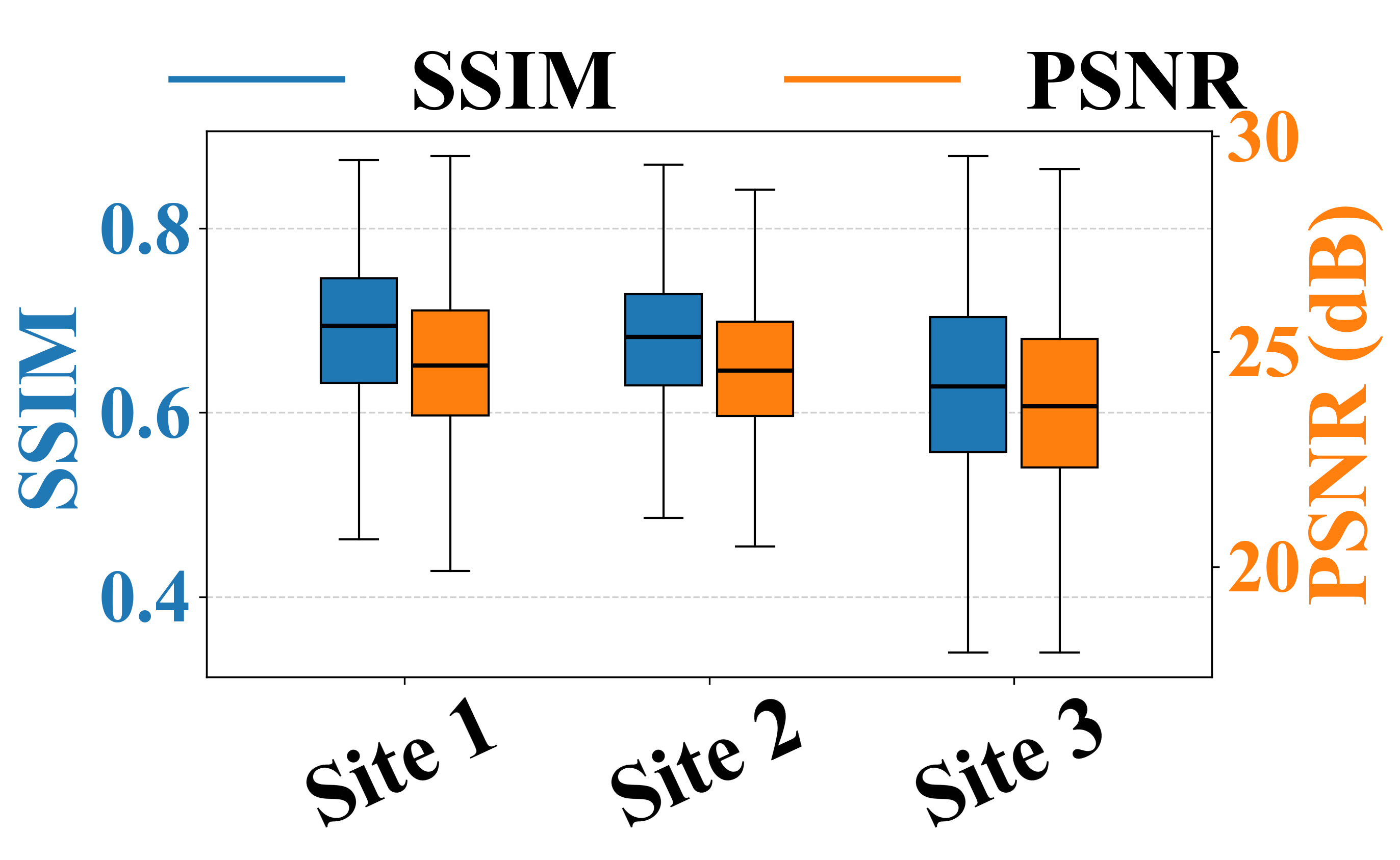}\label{fig:Zonewise_SSIM_PSNR_Boxplot}
    }\hfil
    \vspace{-0.1cm}
    \caption{In-the-wild view adaptation performance.}
    \label{fig:unseenClapBox}\vspace{-0.4cm}
\end{figure}

As shown in \figurename~\ref{fig:unseenActivityBox}, the model consistently achieves high reconstruction fidelity across all three unseen activities, demonstrating strong generalization to motion patterns beyond the training distribution. The \textit{SSIM} values remain in the range of $0.65-0.75$, indicating that the synthesized RD representations preserve fine structural and spatial details. Correspondingly, the \textit{PSNR} values range between $25$~dB and $35$~dB, reflecting low pixel-wise distortion. Overall, these results confirm that the proposed architecture generalizes effectively across \textit{unseen activity categories}.

\subsection{View-Adaptation in Unseen Environments}

To assess the robustness of the view-adaptation module, we evaluated its performance across three distinct deployment environments as introduced in Section~\ref{sec:deployments}. The same view-adaptation network, trained solely on in-lab industrial gestures, was directly applied to these unseen deployment environments without any fine-tuning. As shown in \figurename~\ref{fig:Zonewise_SSIM_PSNR_Boxplot}, the model maintains stable reconstruction quality across all three sites, achieving \textit{SSIM} values in the range of $0.6-0.8$ and \textit{PSNR} values between $22-30$~dB. The outdoor marble processing factory (Site 1) achieves the highest SSIM and PSNR thanks to clear line-of-sight and well-separated workers. The stone-cutting factory (Site 2) shows moderate degradation due to high-speed cutting machinery, whose vibrations interfere with human motion patterns in the RD domain. The indoor construction workspace (Site 3) experiences further challenges from strong multipath and confined geometry, which introduce viewpoint-dependent distortions. Despite these factors, the view-adaptation model reliably preserves overall motion intensity across views.

\subsection{Occupational Pollution Exposure Analysis}
\begin{figure}[h]
    \centering
    \subfigure[Marble processing factory]{
        \includegraphics[width=0.18\textwidth]{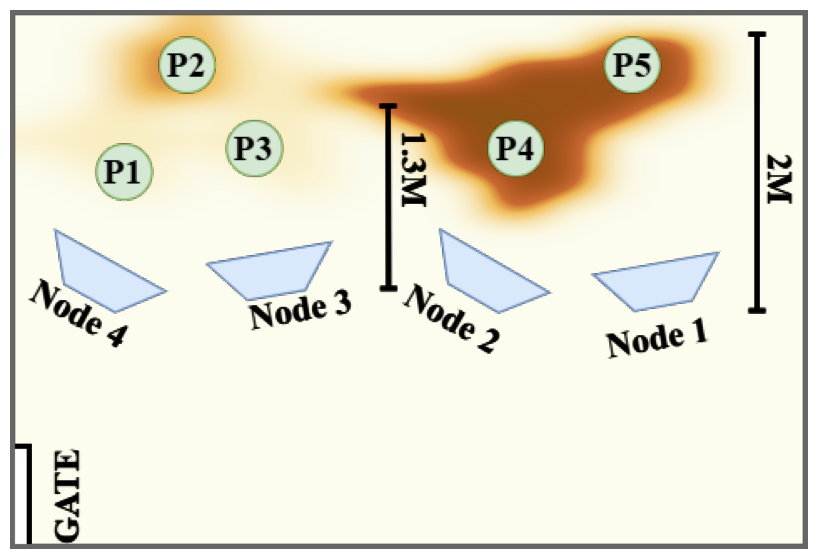}
        \label{fig:pm_marble_outdoor}
    }\hfil
    \subfigure[Stone cutting factory]{
        \includegraphics[width=0.13\textwidth]{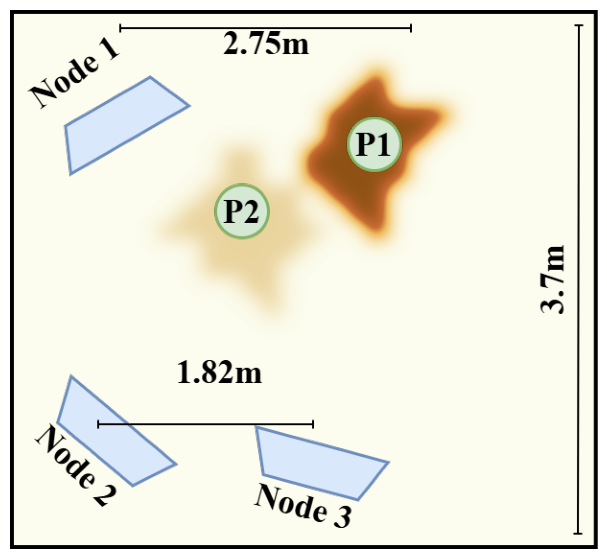}
        \label{fig:pm_khadan}
    }\hfil
    \subfigure[Construction Site]{
        \includegraphics[width=0.12\textwidth]{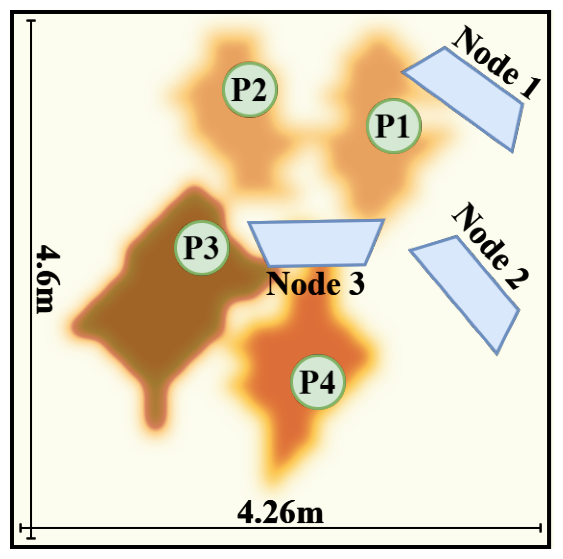}
        \label{fig:pm_indoor}
    }\hfil
    \vspace{-0.2cm}
    \caption{
        Interpolated PM$_{2.5}$ exposure maps across locations.
    }
    \label{fig:pm_exposure_maps}
    \vspace{-0.4cm}
\end{figure}

Table~\ref{tab:pollution_exposure} and Figure~\ref{fig:pm_exposure_maps} present the pollution exposure analysis across different activities, workers, and deployment environments. The results show distinct PM levels across activities and worksites. \emph{Grinding} consistently produces the highest PM levels (PM$_{2.5}$ exceeding 500~$\mu$g/m$^3$), often an order of magnitude higher than those from \emph{polishing} (approximately 100~$\mu$g/m$^3$) or \emph{chipping} (around 350~$\mu$g/m$^3$). Hence, high-friction, abrasive operations like grinding result in significantly higher dust exposure than polishing or standing idle. Figure~\ref{fig:pm_exposure_maps} compares the interpolated PM$_{2.5}$ fields across three representative deployment sites. Due to continuous cutting and high airflow, the outdoor yard shows PM hotspots near grinding areas, whereas the indoor site has more confined PM pockets.

\begin{table}[]
\centering
\scriptsize
\caption{Pollution exposure at various deployment locations.}
\label{tab:pollution_exposure}
\begin{tabular}{|c|c|c|r|r|r|}
\hline
\multicolumn{1}{|c|}{\textbf{Deployment}} & \textbf{Person} & \textbf{Activity}  & \multicolumn{1}{c|}{\textbf{PM$_1$}} & \multicolumn{1}{c|}{\textbf{PM$_{2.5}$}} & \multicolumn{1}{c|}{\textbf{PM$_{10}$}} \\ \hline
\multirow{4}{*}{\textbf{Marble Processing Factory}} 
& P1 & \textbf{Polishing}  & 44 & 106 & 139 \\ \cline{2-6}
& P2 & \textbf{Chipping} & 161 & 360 & 439 \\ \cline{2-6}
& P3 & \textbf{Polishing}  & 44 & 106 & 139 \\ \cline{2-6}
& P4 & \textbf{Grinding}  & 245 & 603 & 703 \\ \cline{2-6}
& P5 & \textbf{Grinding}  & 233 & 563 & 665 \\ \hline
\multirow{2}{*}{\textbf{Stone-Cutting Factory}} 
& P1 & \textbf{Grinding}  & 126 & 544 & 701 \\ \cline{2-6}
& P2 & \textbf{Standing} & 61 & 117 & 125 \\ \hline
\multirow{4}{*}{\textbf{Indoor Construction Site}}
& P1 & \textbf{Sitting}  & 112 & 362 & 415 \\ \cline{2-6}
& P2 & \textbf{Standing} & 118 & 376 & 449 \\ \cline{2-6}
& P3 & \textbf{Grinding}  & 128 & 633 & 898 \\ \cline{2-6}
& P4 & \textbf{Grinding}  & 119 & 717 & 978 \\ \hline
\end{tabular}
\vspace{-0.5cm}
\end{table}

Another critical insight from Table~\ref{tab:pollution_exposure} and Figure~\ref{fig:pm_exposure_maps} is that even passive activities can experience substantial exposure. 
For instance, at the indoor construction site, a worker \emph{sitting} near a grinding area (P1) records a PM$_{2.5}$ level of 362~$\mu$g/m$^3$, comparable to moderate grinding exposure. 
Similarly, \emph{standing} workers near grinding or chipping tasks experience significantly elevated PM readings despite minimal personal activity. 
This proximity-driven exposure underscores the need to jointly consider \textit{activity type}, \textit{spatial position}, and \textit{co-located operations} for occupational health risks. 
 It is worth noting that the global PM map generation incurs an additional delay of approximately 5~s, as the PM sensors operate at 1~Hz and aggregate readings over a 5~s observation window, resulting in an overall latency of about 5.23~s (230 ms of latency in the re-ID pipeline \S\ref{sec:latency}). 

\section{Conclusion}
We presented \ourmethod, a multi-radar privacy-preserving mmWave framework for personalized PM exposure monitoring in harsh industrial settings. Extensive field deployments across marble yards and construction sites demonstrate robust performance, achieving sub-10\,cm localization error and 90.4\% cross-radar re-ID F1-Score. The system further enables fine-grained, per-worker exposure estimation. Overall, \ourmethod demonstrates mmWave sensing as an effective means of monitoring occupational health and safety in environments where optical or wearable methods are not preferred. 

The current \ourmethod framework assumes partial radar overlap and predominantly azimuthal human motion. While effective for the targeted stone-working scenarios, its generalizability may be limited by extreme sensor orientations, sparse node density, or restricted FoV overlap. Environmental factors, such as ghost reflections and dense multi-path interference, also present potential sources of association ambiguity. Resolving these geometric and interference-related challenges defines our primary direction for future research. 

\section*{Ethical Considerations}
All experimental procedures were approved by the Institute Ethical Committee under approval number IIT/SRIC/DEAN/2025/29, dated 25-09-2025 and conducted in accordance with its guidelines. Informed consent was obtained from all participants.

\section*{Acknowledgment}
The authors acknowledge the support of the Department of Science and Technology (DST), Government of India, through the NGP Division under Grant No. NGP/GS-02/Sandip/IIT-K/WB/2023(C), dated 09-01-2024. This work has also been partially supported by the joint DST–US National Science Foundation (NSF) program under Grant No. DST/INT/USA/NSF-DST/NeTS\_Sandip/P4/2024 (G), dated 10-01-2025. In addition, the authors acknowledge support from the Google Academic Research Award for Society-Centric AI for indoor pollution analysis and measurements (DALTON module). 


\balance
  \bibliographystyle{ACM-Reference-Format}
  \bibliography{refs}

\end{document}